\documentclass{article}

\pdfoutput=1
\usepackage{authblk}
\usepackage{fullpage}
\usepackage{amsmath}
\usepackage{booktabs}

\usepackage{caption}
\usepackage{graphicx}

\usepackage[colorlinks=True,
            citecolor=cyan,
            linkcolor=cyan,
            filecolor=cyan,
            urlcolor=cyan]{hyperref}
\usepackage[round, semicolon]{natbib}
\usepackage{multirow}
\usepackage{lscape}

\title{An Extension to the Procedure for Developing Uncertainty-Consistent Shear Wave Velocity Profiles from Inversion of Experimental Surface Wave Dispersion Data}
\author[1]{Joseph P. Vantassel}
\date{}
\author[2]{Brady R. Cox}
\affil[1]{Virginia Polytechnic Institute and State University}
\affil[2]{Utah State University}

\begin{document}

\maketitle

\begin{abstract}
Measurements of shear wave velocity (Vs) with uncertainty are critical for site-specific probabilistic seismic hazard studies. However, rigorously quantifying the uncertainty in Vs over large enough areas and great enough depths remains challenging. In 2021, Vantassel and Cox (i.e., VC21) proposed a procedure for developing suites of Vs profiles from surface wave testing whose uncertainty were consistent with the experimental dispersion data's uncertainty. The VC21 procedure was a significant step forward, however, it requires a full dispersion data matrix to compute inter-wavelength phase velocity correlations. While applicable to many practical cases, VC21 could not be applied to the case where multiple surface wave arrays of different sizes were deployed at a site as a means of developing broadband dispersion data and deeper Vs profiles. In response, this work extends the VC21 procedure using two possible approaches for estimating a full dispersion data matrix. Approach 1 uses a selection of theoretical dispersion curves from an initial, traditional surface wave inversion. Approach 2 estimates the full data matrix by combining pieces of the data matrix obtained from the experimental dispersion measurements. Both approaches are evaluated using two synthetic datasets; one relatively-simple, three-layered model and one more-complex, five-layered model. Approach 1 and Approach 2 were able to reasonably estimate the true correlation matrix and recover uncertainty-consistent Vs profiles similar to the true distribution of Vs. While the uncertainty of the recovered Vs profiles were higher than is often assumed, the engineering proxies computed from those Vs profiles, namely the time averaged shear wave velocity in upper 30 m and the fundamental site period, showed substantially less uncertainty indicating the Vs profiles, while uncertain, are effective at capturing a site's engineering behavior. Both approaches were applied to real data from the Garner Valley Downhole Array (GVDA) site and found to yield better estimates of small-strain site amplification than achieved previously.
\end{abstract}

\pagebreak

\section*{Introduction}

Site-specific measurements of shear wave velocity (Vs), which with mass density ($\rho$) can be used to compute the small-strain shear modulus ($G_0$) (i.e., $G_0=\rho V_s^2$), is a critical input for determining site-specific seismic hazard. While Vs has been traditionally measured using invasive methods, such as downhole, crosshole, or PS suspension logging, non-invasive methods, such as surface wave testing, have been shown to be a cost-effective alternative. In addition, because the use of surface waves inherently involves averaging subsurface material properties (both vertically and laterally), it has been argued that surface wave methods can provide Vs estimates that better represent a site's overall response during earthquake shaking than the point measurements from invasive methods  \citep{griffiths_surface-wave_2016, teague_measured_2018}. Yet, despite the success of surface wave methods as a seismic site characterization tool, the rigorous quantification of uncertainty in the resulting Vs profiles remains challenging. In response, this work proposes to generalize a previously proposed procedure \citep{vantassel_procedure_2021} for developing suites of Vs profiles that rigorously account for the uncertainty in the experimental measurements and non-uniqueness in the inverse problem.

Probabilistic seismic hazard analyses require estimates of uncertainty for all input parameters used to predict site-specific ground shaking. The subsurface Vs structure has been shown to be one of the parameters that significantly impacts ground motion calculations, and the importance of quantifying Vs uncertainty is well understood. In general, probabilistic seismic hazard studies divide the consideration of a site's Vs uncertainty into two parts: epistemic and aleatory. Epistemic uncertainties are a result of lack of knowledge and may be able to be reduced if additional or higher quality information is available. In contrast, aleatoric uncertainties are the result of inherent randomness associated with the selected engineering model that cannot be reduced, but can be quantified. In the context of modeling a site's Vs structure, epistemic uncertainties are accounted for by branches of a logic tree, where each branch contains a different Vs base-case profile and a corresponding confidence weight. These different possible Vs profiles may be the result of different measurements made at a site at different times and/or using different methods. The aleatoric uncertainty associated with each base-case profile for both invasive and surface wave methods has historically been accounted for through the use of Vs randomization \citep{epri_seismic_2012}. Vs randomization approaches, such as that by Toro (\citeyear{toro_probabilistic_1995}) and Passeri et al. (\citeyear{passeri_new_2020}), take the base-case Vs profile as input and modify it in a pseudo-random fashion to develop a suite of uncertain profiles that capture the aleatoric uncertainty believed to be associated with the base-case profile. However, most of the input parameters for Vs randomization cannot be estimated from site-specific observations and therefore must be left to the judgement of the analyst. As a result, several studies have expressed concern over the use of ``blind'' Vs randomization to capture aleatoric uncertainty \citep{griffiths_surface-wave_2016, teague_measured_2018, hallal_comparison_2022} and have highlighted the need for methods to develop suites of Vs profiles for seismic hazard analysis whose uncertainty is site-specific and less directly controlled by (and preferably independent of) analyst judgement.

An alternative to Vs randomization to quantify aleatoric uncertainty for surface wave methods, and the one we advocate for in this paper, is to develop suites of Vs profiles that account for uncertainty in the site-specific surface wave dispersion measurements and the non-uniqueness of the inverse problem. Methods proposed to develop suites of Vs profiles prior to the development of the Vantassel and Cox (\citeyear{vantassel_procedure_2021}) procedure, referred to as the VC21 procedure hereafter, can be roughly divided into two categories. The first category of methods are those that perform a traditional global-search surface wave inversion through many (thousands to hundreds of thousands) of models to develop a population of Vs profiles that are then sampled to produce a manageable number of profiles for subsequent analysis. These methods differ according to their choice of surface wave inversion algorithm \citep{wathelet_surface-wave_2004, foti_non-uniqueness_2009, hollender_characterization_2018}, the number of models considered in the inversion \citep{wathelet_surface-wave_2004, wathelet_improved_2008, teague_development_2015, hollender_characterization_2018}, and the approach to Vs profile sampling \citep{foti_non-uniqueness_2009, di_giulio_exploring_2012, teague_development_2015, deschenes_development_2018}. Nonetheless, all use a very similar methodology and can be readily implemented using all existing global-search inversion algorithms. However, as shown in Vantassel and Cox (\citeyear{vantassel_procedure_2021}), these methods do not rigorously capture the uncertainty in the site-specific surface wave dispersion measurements and the non-uniqueness of the inverse problem. They should therefore be considered approximate with a tendency to underestimate uncertainty in a site's Vs structure. The second category of methods are those that use a Bayesian framework for performing the surface wave inversion and quantifying uncertainty \citep{molnar_bayesian_2010}. The development of a Bayesian approach to surface wave inversion is substantially more involved than the simplified methods aforementioned. Bayesian methods require defining a likelihood function that incorporates assumptions regarding the distribution of data uncertainty, estimating a covariance structure for the experimental data, quantifying priors for each inversion parameter considering existing knowledge, and implementing advanced sampling tools to numerically evaluate the integrals that define the most likely model and associated uncertainty distributions \citep{molnar_bayesian_2010, dettmer_trans-dimensional_2012, gosselin_gradient-based_2017}. The definition of these components is not trivial and each decision in the development process requires careful consideration of the associated statistical implications. Yet, these methods have been shown to be statistically rigorous in their quantification of uncertainties and able to be used for subsequent engineering analyses \citep{molnar_uncertainty_2013, gosselin_probabilistic_2018}. In summary, the first category of methods are relatively-simple and can be implemented with any global-search inversion algorithm but tend to underestimate Vs uncertainty. In contrast, the second category of methods are rigorous in their consideration of Vs uncertainty but are substantially more complex to implement and require a thorough understanding of Bayesian statistics. As an alternative, the VC21 procedure provides a rigorous means of quantifying uncertainty in surface wave inversion in an easy-to-implement, inversion algorithm agnostic framework that can be implemented without advanced statistical knowledge.

While the reader is referred to the original study, we briefly reiterate the key points of the VC21 procedure for developing uncertainty-consistent Vs profile from surface wave dispersion measurements. First, the uncertainty in the experimental dispersion data must be quantified. The quantification of experimental dispersion data uncertainty has been discussed extensively in the literature \citep{lai_propagation_2005, foti_non-uniqueness_2009, cox_surface_2011} with Vantassel and Cox (\citeyear{vantassel_swprocess_2022}) detailing a systematic workflow that integrates this prior work with a freely-available, open-source software tool called \emph{swprocess} \citep{vantassel_jpvantasselswprocess_2021}. But, in short, the quantification of experimental dispersion data uncertainty involves deploying surface wave arrays at multiple locations and across multiple scales using active-source and passive-wavefield methods to make many observations of the site's experimental dispersion data. The authors note that the quantity of observations necessary to robustly characterize a site's experimental dispersion data uncertainty remains unsettled, in particular, there are open questions regarding to what extent metrological and seasonal variations influence a site's apparent uncertainty and how those should be best quantified. However, as a practical solution while research continues in this area the authors recommend collecting as much surface wave data as time permits, using different arrays over different spatial scales, to characterize the site and its uncertainty as robustly as possible. Second, after data collection and processing the observations of the site's experimental dispersion data are assembled in a data matrix. The dispersion data matrix is a way of organizing the experimental dispersion measurements where each row is a corresponding observation (e.g., phase velocity estimates extracted from a single shot gather or microtremor time window) and each column is a corresponding variable (e.g., wavelengths where those observations were made). Third, the dispersion data matrix, which is assumed to be full in the VC21 procedure, is then used to compute the mean vector and covariance matrix between the selected variables (i.e., set of wavelengths where phase velocities were measured). The experimental dispersion data are then modeled as a multivariate normal distribution using the associated mean vector and covariance matrix. Fourth, realizations from the multivariate normal distribution are drawn and inverted until the theoretical dispersion data associated with the inverted Vs profiles appropriately captures the experimental dispersion data uncertainty. The associated suite of Vs profiles shown to be consistent with the uncertainty of the experimental data are termed uncertainty-consistent Vs profiles. This process is repeated for multiple inversion parameterizations, where each parameterization is defined with a different number of layers to account for the epistemic uncertainty regarding the site's true number of layers. If site-specific data is available, epistemic uncertainty in the site's number of layers can be reduced and the analyst may be able to consider fewer parameterizations and correspondingly fewer logic tree branches. Importantly, we note previous work has attempted to address the challenge of parametrization uncertainty. These efforts broadly include the use of statistical methods for combining data fit with parametrization complexity into a quantitative measure to highlight the simplest model that fits the data \citep{di_giulio_exploring_2012, molnar_bayesian_2010} as well as transdimensional methods that treat the model parametrization as a quantity to be optimized \citep{bodin_transdimensional_2012, dettmer_trans-dimensional_2012}. Work in these areas are important and on-going, but herein we choose to invert the experimental dispersion data using multiple parameterizations as any given inversion parameterization implies knowledge about the site's subsurface structure that, without site specific data, is unknown. By using methods to select only a single parameterization, despite it being the simplest or the most optimized, we have concerns that this may overlook features (e.g., strong impedance contrasts) that can have significant impact on a site's seismic amplification. In summary, the VC21 procedure proposed to replace randomized base-case Vs profiles when performing surface wave inversion with suites of uncertainty-consistent Vs profiles where each suite is generated from an inversion parameterizations with a different numbers of layers. The authors note that while there may be some similarity between the VC21 procedure and that of the bootstrap from statistics \citep{efron_bootstrap_1979}, they are different in their assumptions. The bootstrap assumes independence between samples whereas the VC21 procedure assumes correlation, the bootstrap samples the original data with replacement whereas the VC21 procedure samples from a statistical model of the data, and the bootstrap is distribution independent whereas the VC21 procedure assumes a multivariate normal distribution.

\begin{figure}[t!]
    \centering
	\includegraphics[width=1.0\textwidth]{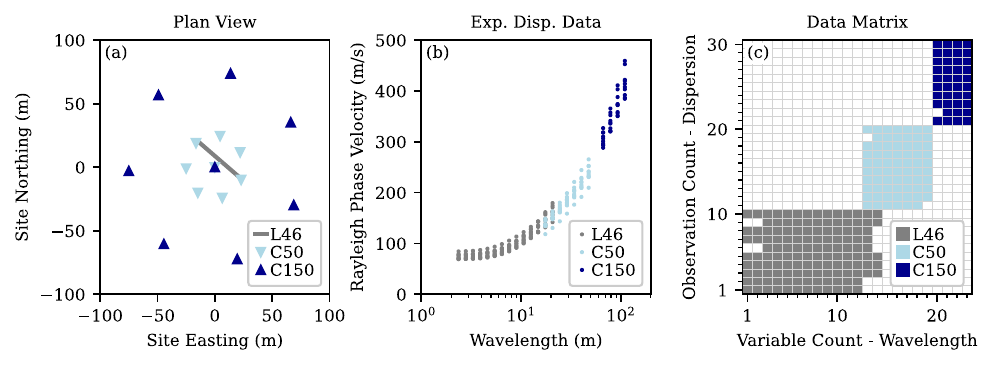}
	\caption{Illustration of the incomplete data matrix problem when deploying (a) surface wave arrays of different size to (b) measure experimental dispersion data over different wavelength ranges (c) resulting in a data matrix that is largely incomplete. Panel (a) shows a plan view of three surface wave arrays, one linear active-source array (L46) and two circular passive-wavefield arrays (C50 and C150), deployed at a hypothetical site. Panel (b) shows experimental dispersion data extracted from each array with the smaller arrays measuring shorter wavelengths and the larger arrays measuring longer wavelengths. Panel (c) presents the incomplete dispersion data matrix where the data from each array is indicated by color and no-color indicates missing data. Note that data shown is hypothetical and in general more than 10 observations would be made per variable to develop robust statistics.}
	\label{fig:1}
\end{figure}

The VC21 procedure outlined above works well when the data matrix is full or nearly full. In the case of the latter, the nearly full data matrix can be completed using existing, variance-preserving imputation techniques from traditional statistics \citep{enders_applied_2022}. A common example where the data matrix is full, or nearly full, includes the case where one or more surface wave arrays of similar size are deployed at a site and used to measure wavelengths over a similar range, as shown in the original VC21 paper. Another example includes the case where a single surface wave array excited by different active-sources observes multiple surface wave modes, as presented in Vantassel et al. (\citeyear{vantassel_extracting_2022}). However, the VC21 procedure does not account for the cases where surface wave arrays of different sizes are deployed at the same site as a means to develop deeper Vs profiles and where each array primarily observes a unique wavelength range. This is very commonly done when combining active-source and passive-wavefield surface wave methods. To more-clearly illustrate the challenge to the VC21 procedure for this situation, the case in question is illustrated in Figure \ref{fig:1}. Figure \ref{fig:1}a shows a plan view of a hypothetical site where three surface wave arrays of different sizes (i.e., L46, C50, and C150) are deployed. The three arrays are denoted by a letter indicated their shape, linear (L) or circular (C), and their corresponding length/diameter in meters. For example, the L46 array is a common array used for active-source multichannel analysis of surface waves (MASW) testing, where 24 geophones are deployed at a constant 2-m spacing, while the C50 or C150 arrays are commonly used for microtremor array measurements (MAM), where eight seismometers are deployed to record ambient noise. Each array is used to observe a different segment of the site's surface wave dispersion data, as shown in Figure \ref{fig:1}b, with the larger arrays observing the longer wavelengths. Using all three arrays allows for a wide range of wavelengths to be observed, enabling a more complete quantification of the site's dispersion characteristics. Finally, these observations of the site's dispersion characteristics can be organized into a data matrix, illustrated in Figure \ref{fig:1}c, where the rows are observations of the phase velocity dispersive trend and the columns are the discrete wavelengths where the experimental data have been sampled. The colored boxes indicate which array has observed the Rayleigh phase velocity value; the absence of color indicates data are missing. Note that it is not uncommon to have experimental dispersion data not-observed near the upper and lower wavelength limit, these are also shown in Figure \ref{fig:1}c. Figure \ref{fig:1}c demonstrates the majority (68\% for this example) of the data matrix is empty, preventing the VC21 uncertainty-consistent procedure from being applied directly.

This work seeks to understand how to best estimate the experimental dispersion data's correlation matrix when multiple surface wave arrays of different size are deployed and the dispersion data matrix is incomplete. In particular, the work proposes and evaluates two alternatives: Approach 1 and Approach 2. Approach 1 is more-time consuming than Approach 2, but can still be used when the raw experimental dispersion data from each array is unavailable and only the overall mean and standard deviation of the data is provided. Approach 2 is faster than Approach 1, but does require the raw experimental dispersion data from each individual array to be available. After both approaches are presented in detail, they are applied to two synthetic examples where the true correlation matrix and Vs profile distribution are available for comparison. Following the synthetic examples, both approaches are applied to a real dataset from the Garner Valley Downhole Array (GVDA) site in southern California. The uncertainty-consistent profiles are compared to measurements made previously at the GVDA site in terms of Vs and their ability to predict observed small-strain site effects.

\section*{Estimating the Correlation Matrix When the Data Matrix is Incomplete}

Two approaches are proposed for estimating the inter-wavelength correlation matrix when the data matrix is incomplete. These approaches are presented in the following sections. First, however, we briefly explain in an intuitive fashion why the assumption of independence between wavelengths (i.e., no inter-wavelength correlation) will result in an underestimation of Vs uncertainty. In the VC21 procedure, assuming a suitably rigorous inversion algorithm is used, the mapping between the experimental dispersion data uncertainty and the uncertainty-consistent Vs profiles relies primarily on the quality of the realizations of the experimental dispersion data. In the case where no correlation is assumed between wavelengths the realizations of the experimental dispersion data are highly variable from wavelength to wavelength, as expected for independent random variables. However, when inverted these highly erratic trends cannot be captured by a relatively simple inversion parameterizations, and only the average trend of the data is captured (i.e., regression to the mean). As a result, the uncertainty implied by the theoretical dispersion curves resulting from the uncertainty-consistent inversion, and by extension the associated uncertainty-consistent Vs profiles, will underestimate the experimental dispersion data uncertainty. These results are consistent with our intuition that assuming correlated variables are independent will results in underestimation of their implied uncertainty. As such, to determine uncertainty-consistent Vs profiles using the VC21 procedure that properly account for the experimental dispersion data uncertainty it is paramount that the inter-wavelength correlation be estimated.

\subsection*{Approach 1}

\begin{figure}[t!]
    \centering
	\includegraphics[width=1.0\textwidth]{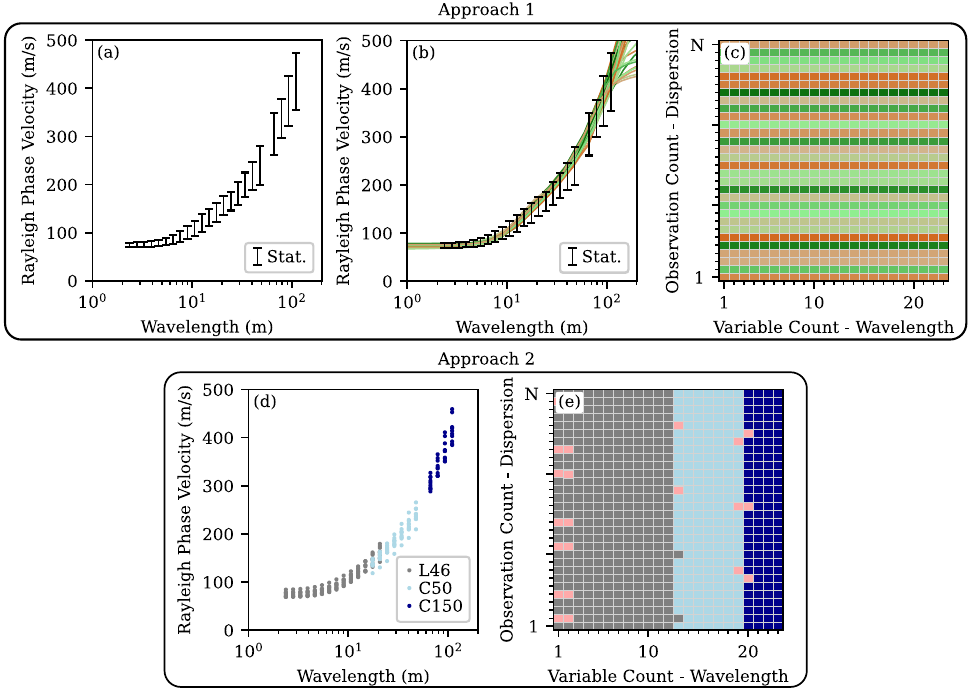}
	\caption{Two proposed approaches to solving the incomplete dispersion data matrix problem. Approach 1 uses the (a) statistical representation of the experimental dispersion data to perform a global-search inversion to recover (b) N observations of the theoretical dispersion curves that approximately fit the statistical representation of the experimental dispersion data, and create a (c) surrogate data matrix for computing the inter-wavelength correlation structure. Note that each theoretical dispersion curve from (b) maps to a row of the surrogate data matrix in (c), providing a phase velocity observation at each wavelength. To clearly illustrate this correspondence the theoretical dispersion curves in (b) are presented with the same color as their corresponding data matrix row in (c). Approach 2 uses every combination of the (d) raw experimental dispersion data to develop (e) a full (or near full) surrogate data matrix. If the data matrix is not full, as shown here, variance preserving imputation is used for the missing values (pink colored entries). The full surrogate data matrix is then used to estimate the inter-wavelength correlation structure. }
	\label{fig:2}
\end{figure}

The first approach uses theoretical dispersion curves from an initial, non-uncertainty-consistent surface wave inversion using a global-search algorithm to develop an estimate of the full dispersion data matrix. Approach 1 starts with the statistical representation of the experimental dispersion data. This data, illustrated in Figure \ref{fig:2}a, is effectively a vector of means and a corresponding vector of standard deviations, one per desired wavelength (or frequency), calculated from the raw dispersion data extracted from each array (i.e., data presented in Figure \ref{fig:1}b). Note that Approach 1 does not require the underlying raw experimental dispersion data and therefore can be applied to historical datasets even when the raw data from each array is no longer available. Next, the statistical representation of the experimental dispersion data is inverted using a global-search algorithm. Notably, the objective of this inversion is not to recover the single, best fit to the experimental dispersion data, but rather to find a diverse set of plausible models whose theoretical dispersion curves fit the experimental dispersion data reasonably well. This can be done using any global-search inversion algorithm by enforcing the misfit (or error function) to some minimum value. By setting a minimum misfit the global-search algorithm is limited in its ability to narrow in on the exact solution, thereby forcing it to find many plausible solutions. For this study, we will use the Neighborhood Algorithm \citep{sambridge_geophysical_1999} as implemented by Wathelet (\citeyear{wathelet_array_2005}, \citeyear{wathelet_improved_2008}) in the software Dinver released as part of the Geopsy suite \citep{wathelet_geopsy_2020} as our surface wave inversion engine. We note that, while not presented in detail here, we explored using multiple different minimum misfit thresholds in Dinver between zero and one, but ultimately found that for the problems considered, the correlation matrix estimates were similar, with the minimum misfit of one results tending to be the most similar to the true correlation matrix. Therefore, we recommend the use a minimum misfit of one when using Dinver to implement this procedure and will use this value throughout the remainder of this work. We note that performing a minimum misfit inversion can converge more-slowly than a standard inversion (i.e., a minimum misfit of zero inversion), however we were able to achieve inversion results with a misfit less than one after the consideration of 60,000 trial models with convergence to a misfit of one typically occurring after 20,000 trial models. After the initial inversion is complete, a select number of the N best models, we use 100 in this study, are extracted from the inversion results. The theoretical dispersion curves from these models are then compared to the experimental dispersion data (see Figure \ref{fig:2}b) and sampled at each wavelength of the experimental dispersion data to populate the surrogate data matrix (see Figure \ref{fig:2}c). However, from results presented in VC21, we know the surrogate data matrix will not robustly capture the mean and uncertainty of the experimental dispersion data. This is shown clearly with the theoretical dispersion curves failing to ``fill up'' the error bars of the experimental dispersion data (Figure \ref{fig:2}b, specifically between the wavelengths of 10 and 50 m). As such we cannot use it directly for implementing the VC21 procedure. Instead, we only consider the inter-wavelength correlation matrix as reliable. To ascertain the mean vector and covariance matrix necessary for implementing the VC21 procedure we use the mean values calculated from the raw experimental dispersion data and compute the covariance matrix by scaling the inter-wavelength correlation matrix from the surrogate data matrix by the standard deviation values from the raw experimental dispersion data. Approach 1 then results in a statistical model whose mean and standard deviation of phase velocity at each wavelength is the same as would have been determined from the experimental dispersion data directly (i.e., if Approach 1 had not been used) and an estimate of inter-wavelength correlation from the physics-informed surrogate data matrix. Together these define a mean vector and covariance matrix that can be used with the VC21 procedure for developing uncertainty-consistent Vs profiles.

\subsection*{Approach 2}

The second approach combines the available, band-limited experimental dispersion data from each array to develop an estimate of the full data matrix. Figure \ref{fig:2}d illustrates that Approach 2 does not impose any restrictions on the form of the raw experimental dispersion data. For example, the data from different arrays may overlap (e.g., at approximately 20 m wavelength) or have gaps (e.g., approximately 60 m wavelength). Approach 2 combines the observed experimental dispersion data to develop pseudo-observations where one segment is selected from each array's wavelength range to form the full surrogate data matrix. For the hypothetical case shown in Figure \ref{fig:2}e, one observation would be selected from each of the arrays (i.e., L46, C50, C150) to form one row of the surrogate data matrix. Recall that observations for the active-source arrays may correspond to different shot locations, while observations from passive arrays might correspond to different time windows. The first row of the data matrix becomes the first observation from each array, the second row the first observation from the first two arrays and the second from the third, and so forth until the last row is defined by the last observation from each array. As such, for a given set of experimental dispersion data the estimate of the data matrix is fully defined. In the case of dispersion data overlap (e.g., between the L46 and C50 array at approximately 20 m wavelength), a decision must be made regarding which of the data is to be selected or if the data should be statistically combined. In this study we opt to select the data observed by the larger array, because the longest wavelengths observed by an array are generally less reliable than the shortest wavelengths observed (e.g., Lai et al., \citeyear{lai_propagation_2005}; Garofalo et al., \citeyear{garofalo_interpacific_2016}). The process of selecting observations from each array are repeated until all N combinations have been used to populate the surrogate data matrix. If many arrays are deployed at a site with overlapping wavelength ranges, the data from these arrays can be combined to reduce the number of combinations to be computed. If the resulting surrogate data matrix is incomplete (like the one shown in Figure \ref{fig:2}e) variance-preserving imputation is used, where random values informed by the known dispersion data's mean and standard deviation are used to ``fill in'' the missing data values (shown with pink-colored entries in Figure \ref{fig:2}e). Importantly, Approach 2, unlike Approach 1, requires the raw experimental dispersion data from each array to be available, limiting its applicability to historical datasets. However, Approach 2 is faster to perform than Approach 1, because it does not rely on an initial, non-uncertainty-consistent surface wave inversion to develop an estimate of the full data matrix. In quantitative terms, on a typical computer with an experienced analyst Approach 2 will take approximately 5 minutes to estimate the full data matrix whereas Approach 1 will take approximately 30 minutes. We evaluate the effectiveness of these two approaches in the following sections.

\section*{Synthetic Data Examples}

The effectiveness of Approach 1 and Approach 2 are evaluated using two synthetic datasets. The first dataset corresponds to a relatively-simple three-layered model and the second to a more-complex five-layered model. The use of synthetic models allow for the estimated correlation matrix and uncertainty-consistent Vs profiles from Approaches 1 and 2 to be compared to the true correlation matrix and true distribution of Vs.

\subsection*{Simple Synthetic Dataset: Three-Layered Model}

\begin{figure}[t!]
    \centering
	\includegraphics[width=1.0\textwidth]{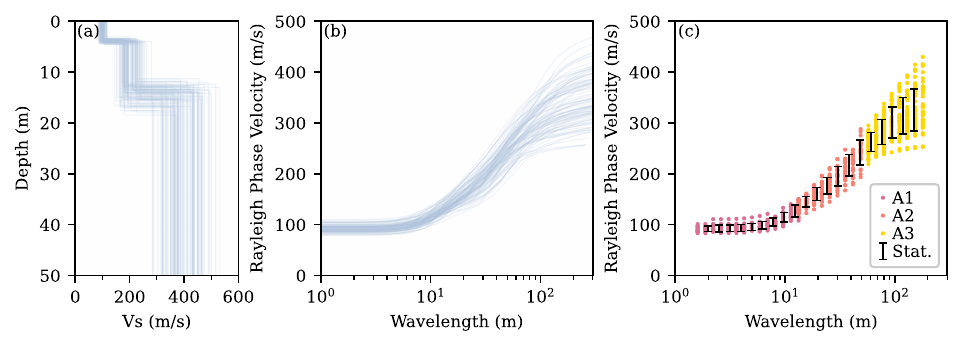}
	\caption{Creation of a synthetic experimental dispersion dataset for a \textbf{three-layered model} by first (a) generating 100 hypothetical shear wave velocity (Vs) profiles, (b) computing their corresponding theoretical dispersion data, and (c) defining the experimental dispersion data (i.e., sampled theoretical dispersion curves) over specific wavelength ranges as if they were measured by three arrays (A1, A2, and A3) of different size. The statistical representation of the synthetic experimental dispersion data is also presented in (c) with $\pm$ one standard deviation error bars.}
	\label{fig:3}
\end{figure}

Approaches 1 and 2 are first evaluated on a three-layered synthetic model. To develop experimental dispersion data with associated uncertainty that correspond to a known distribution of Vs with depth, we first develop a statistical model for the Vs profiles. We assume the thickness and Vs of each layer can be modeled as a multivariate-lognormal distribution, and define the mean thickness and Vs of each layer to follow Profile 5 from the surface wave inversion benchmark dataset developed by Vantassel and Cox (\citeyear{vantassel_surface_2020}). We define the lognormal standard deviation of the Vs and thickness of each layer as well as the interparameter correlation structure such that they produce realistic realizations. Our two criteria for realism are: (1) the Vs profiles exhibit typical levels of uncertainty which we consider a lognormal standard deviation of Vs ($\sigma_{ln,Vs}$) with depth of approximately 0.15 \citep{stewart_guidelines_2014} and (2) the implied experimental dispersion data exhibit typical levels of uncertainty which we consider a coefficient of variation of 0.05 at short wavelengths increasing up to 0.15 at long wavelengths \citep{garofalo_interpacific_2016}. The exponentiated lognormal median ($\mu_{ln}^*$) and lognormal standard deviation ($\sigma_{ln}$) of each layer's Vs and thickness (H) for the three-layered model are provided in Table \ref{table:1}. With the geostatistical model of the Vs profiles defined, we generate 100 Vs profile realizations, as shown in Figure  \ref{fig:3}a. Note these profiles represent the full statistical distribution of models with many clustered near the model's mean profiles and fewer models distant from the mean. To be able to compute the theoretical dispersion curves, we define the compression wave velocity (Vp) of each layer to be twice Vs (i.e., defining Poisson's ratio of 0.33) and define the mass density ($\rho$) to be 2000 $kg/m^3$. We note that while more sophisticated means of defining Vp and $\rho$ could be used (e.g., from correlations from the literature) we do not do so here, as it is known that surface wave dispersion is not highly sensitive to Vp and $\rho$ \citep{wathelet_surface-wave_2004}. With the H, Vs, Vp, and $\rho$ defined for each realization, the associated theoretical dispersion curves can be computed (one per model), as shown in Figure \ref{fig:3}b. Finally, the theoretical dispersion curves are split into three groups: 20 realizations to group A1, 30 realizations to group A2, and 50 realizations to group A3. Groups A1, A2, and A3 are truncated to the wavelength ranges of 1 – 14 m, 13 – 50 m, and 50 – 200 m, to produce data that appears as if it was acquired by three surface wave arrays of different size. The simulated experimental dispersion data and corresponding $\pm$ one standard deviation dispersion statistics are shown in Figure \ref{fig:3}c.

\begin{table}
\centering
\caption{Exponentiated lognormal median ($\mu_{ln}^*$) and lognormal standard deviation ($\sigma_{ln}$) of the parameters [i.e., layer-wise shear wave velocity (Vs) and thickness (H)] defining the inversion results of the synthetic \textbf{three-layered model}. Results are presented for the true distribution and for the distributions resulting from both Approach 1 and Approach 2 after the 1st and 2nd uncertainty-consistent inversion.}
\label{table:1}
\resizebox{\textwidth}{!}{%
\begin{tabular}{@{}ccccccccccc@{}}
\toprule
          &                   &                   & \multicolumn{4}{c}{Approach 1}                                                  & \multicolumn{4}{c}{Approach 2}                                                  \\
          & \multicolumn{2}{c}{True Distribution} & \multicolumn{2}{c}{Post 1st Inversion} & \multicolumn{2}{c}{Post 2nd Inversion} & \multicolumn{2}{c}{Post 1st Inversion} & \multicolumn{2}{c}{Post 2nd Inversion} \\
Parameter & $\mu_{ln}^*$      & $\sigma_{ln}$     & $\mu_{ln}^*$      & $\sigma_{ln}$      & $\mu_{ln}^*$      & $\sigma_{ln}$      & $\mu_{ln}^*$      & $\sigma_{ln}$      & $\mu_{ln}^*$      & $\sigma_{ln}$      \\ \midrule
Vs1 (m/s) & 100               & 0.07              & 97                & 0.07               & 97                & 0.07               & 99                & 0.07               & 97                & 0.07               \\
Vs2 (m/s) & 200               & 0.1               & 189               & 0.27               & 195               & 0.25               & 197               & 0.21               & 203               & 0.22               \\
Vs3 (m/s) & 400               & 0.15              & 391               & 0.19               & 386               & 0.19               & 386               & 0.16               & 382               & 0.15               \\
H1 (m)    & 4                 & 0.05              & 3.8               & 0.27               & 3.9               & 0.25               & 3.9               & 0.17               & 4                 & 0.13               \\
H2 (m)    & 10                & 0.1               & 11                & 0.72               & 10                & 0.74               & 11                & 0.58               & 11                & 0.59               \\ \bottomrule
\end{tabular}%
}
\end{table}

\begin{figure}[t!]
    \centering
	\includegraphics[width=1.0\textwidth]{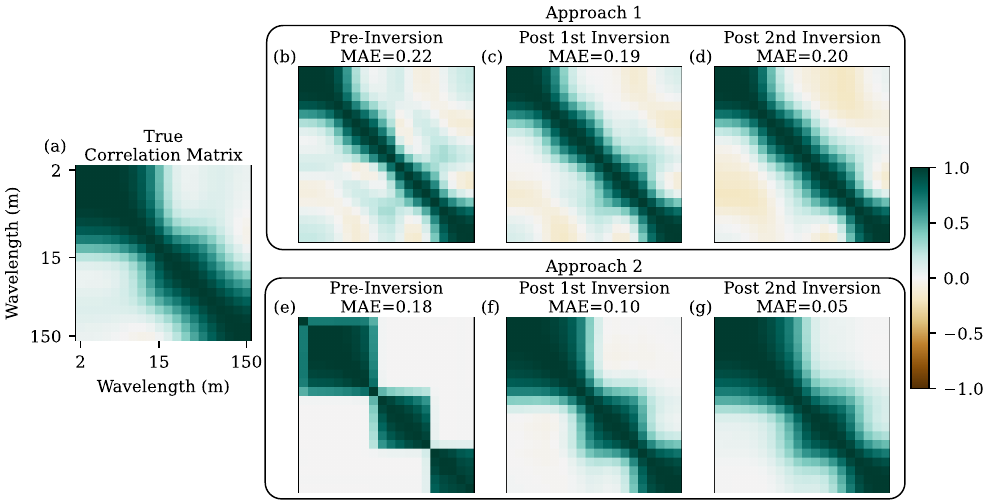}
	\caption{The correlation matrices estimated using Approach 1 and Approach 2 for the synthetic experimental dispersion dataset derived from a \textbf{three-layered model} compared to the (a) true correlation matrix. The correlation matrices for Approach 1 and 2 are shown (b \& e) before the 1st uncertainty-consistent inversion, (c \& f) after the 1st uncertainty-consistent inversion, and (d \& g) after the 2nd uncertainty-consistent inversion. The mean absolute error (MAE) between the estimated correlation matrix and the true correlation matrix is listed above each panel. All correlation matrices are plotted with the same orientation and scale as (a).}
	\label{fig:4}
\end{figure}

Approaches 1 and 2 were then applied to the experimental dispersion data to develop the surrogate data matrix to compute the inter-wavelength correlation structure. For Approach 1, we performed a minimum misfit of one (i.e., M1) inversion of the dispersion statistics computed from the synthetic dispersion data statistics (recall Figure \ref{fig:3}c) using the Dinver module of the open-source Geopsy software. For the inversion, we used a Layering by Number (LN) of 3 (i.e., LN=3) parameterization \citep{vantassel_swinvert_2021}, which was consistent with the true model, and did not consider multiple parameterizations as a simplification. The inversion was performed using three trials, with 60,000 models searched per trial. The 180,000 models were combined and the 100 lowest misfit models (all of which attained the minimum misfit of one) were selected for defining the surrogate data matrix. The resulting inter-wavelength correlation structure is presented in Figure \ref{fig:4}, with Figure \ref{fig:4}a showing the true correlation structure and Figure \ref{fig:4}b showing the correlation structure estimated using Approach 1. The correlation matrix presented in Figure \ref{fig:4}b, which we will refer to as the pre-inversion correlation matrix for clarity, was then used with the synthetic experimental dispersion statistics in an uncertainty-consistent inversion following VC21. The uncertainty-consistent inversion used the LN=3 parameterization developed for performing the initial M1 inversion to invert 250 realizations of the experimental dispersion data. The theoretical dispersion curves implied by the 250 inverted ground models were used to compute the post-inversion inter-wavelength correlation structure, shown in Figure \ref{fig:4}c. A visual comparison of Figure \ref{fig:4}c and \ref{fig:4}b with \ref{fig:4}a indicates that the post-inversion correlation matrix in Figure \ref{fig:4}c may provide a slightly better estimate of the true correlation matrix than the pre-inversion correlation in Figure \ref{fig:4}b. Therefore, it was of interest to repeat the uncertainty-consistent inversion using the post-inversion inter-wavelength correlation structure (i.e., Figure \ref{fig:4}c) to see if additional improvement could be obtained. Figure \ref{fig:4}d presents the inter-wavelength correlation structure after the 2nd uncertainty-consistent inversion, with improvement toward the true solution being less apparent. To quantify the quality of the three inter-wavelength correlation matrices, the pixel-by-pixel mean absolute error (MAE) was computed between each estimate and the true correlation matrix and is listed above each panel in Figure \ref{fig:4}. The MAE corroborates the qualitative observations made previously, in particular, that a slight improvement was obtained after the 1st uncertainty-consistent inversion with no improvement after the 2nd uncertainty-consistent inversion. A similar process is repeated for Approach 2, with the results shown in Figure \ref{fig:4}e, \ref{fig:4}f, and \ref{fig:4}g. In this case, however, notable improvement to the estimates of the correlation matrix were observed after both the 1st and 2nd uncertainty-consistent inversion. In conclusion, Figure \ref{fig:4} illustrates that inter-wavelength correlation structure can be reasonably approximated using both Approach 1 and 2 for the three-layered synthetic model, with Approach 2 providing better MAE values.

\begin{figure}[t!]
    \centering
	\includegraphics[width=0.5\textwidth]{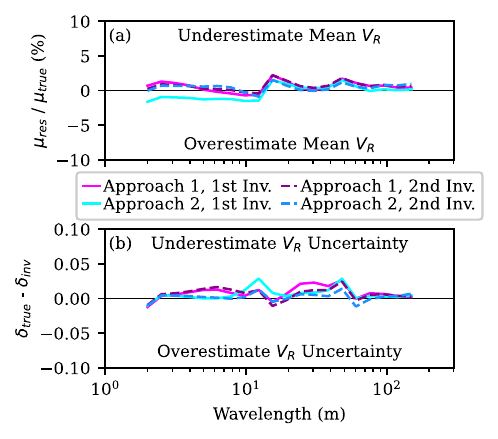}
	\caption{Evaluation of the uncertainty-consistent inversions of the synthetic experimental dispersion data derived from a \textbf{three-layered model} in terms of: (a) the phase velocity residual mean ($\mu_{res}$)[i.e., the difference between the mean of the true experimental dispersion data ($\mu_{true}$) and the mean of the inverted theoretical dispersion curves ($\mu_{inv}$)] normalized by $\mu_{true}$ and expressed in percent, (b) the phase velocity residual coefficient of variation [i.e., the difference between the coefficient of variation of the true experimental dispersion data ($\delta_{true}$) and the coefficient of variation of the inverted theoretical dispersion curves ($\delta_{inv}$)]. Results are shown for both Approach 1 and Approach 2 after the 1st (solid) and 2nd (dashed) uncertainty-consistent inversion.}
	\label{fig:5}
\end{figure}

In addition to the correlation structure, it is important to evaluate the implied mean and standard deviation of the phase velocity after each of the uncertainty-consistent inversions aforementioned to ensure accurate quantification of the full multivariate normal distribution. Figure \ref{fig:5}a presents the difference between the true and inverted mean phase velocity (i.e., $\mu_{true}$ and $\mu_{inv}$, respectively) normalized by $\mu_{true}$ at each wavelength. For convenience we refer to the difference between the true and inverted mean phase velocity as the residual phase velocity ($\mu_{res}$). Figure \ref{fig:5}b presents the difference between the true and inverted coefficient of variation (i.e., $\delta_{true}$ and $\delta_{inv}$, respectively) at each wavelength. These data are presented for all four uncertainty-consistent inversions aforementioned. All four uncertainty-consistent inversions are observed to reside within approximately $\pm$ 1\% of the true mean phase velocity over the majority of the wavelength bandwidth, with the exception of slightly larger deviations at approximately 15 m and 50 m. The wavelengths of 15 m and 50 m coincide approximately with where the synthetic experimental dispersion data was divided between hypothetical arrays and, therefore, the larger than average deviations from the mean are believed to be the result of the associated loss of correlation between adjacent wavelengths. These deviations, which are shown in the following five-layered and real data examples but to a lesser degree for reasons discussed later, serve as a reminder of the importance of incorporating inter-wavelength correlation to the accurate quantification of uncertainty. Nonetheless, Figure \ref{fig:5}a indicates that all four uncertainty-consistent inversions were able to accurately capture the mean trend of the synthetic experimental dispersion data. For the residual coefficient of variation presented in Figure \ref{fig:5}b, all inversion are within $\pm$ 0.03, with the largest differences observed at approximately 11 m and 50 m, again in close proximity to the dividing wavelengths. Of particular note in Figure \ref{fig:5}b is the slight bias that is observed in the residual coefficient of variation, with the residual indicating slight underestimation of Rayleigh wave uncertainty across the wavelength range. We attribute this underestimation to the three-layered parameterization strongly constraining the uncertainty-consistent inversion, such that the more-complex realizations of the simulated experimental dispersion data statistics (that imply a more-complex subsurface structure) may not have been able to be well-fit with three layers, resulting in a corresponding variance reduction. This attribution is based on results to be presented later when more-complex parameterization are used and this bias in the residual coefficient of variation is not observed. Nonetheless, the results presented in Figure \ref{fig:5} indicate that uncertainty-consistent inversions were all able to capture the mean and variance of the experimental dispersion data within 3\% or less, indicating the uncertainty-consistent inversions were successful. In addition, while Figure \ref{fig:4} indicated that by repeating the uncertainty-consistent inversion it was possible to improve the quality (i.e., reduce the MAE) of the inter-wavelength correlation matrix, no such improvement is observed in terms of the capture of the mean and variance of the experimental data. Therefore, as each uncertainty-consistent inversion is time-consuming, these results hint that the additional inversion may not be providing a benefit sufficient to outweigh the associated cost.

\begin{figure}[t!]
    \centering
	\includegraphics[width=1.0\textwidth]{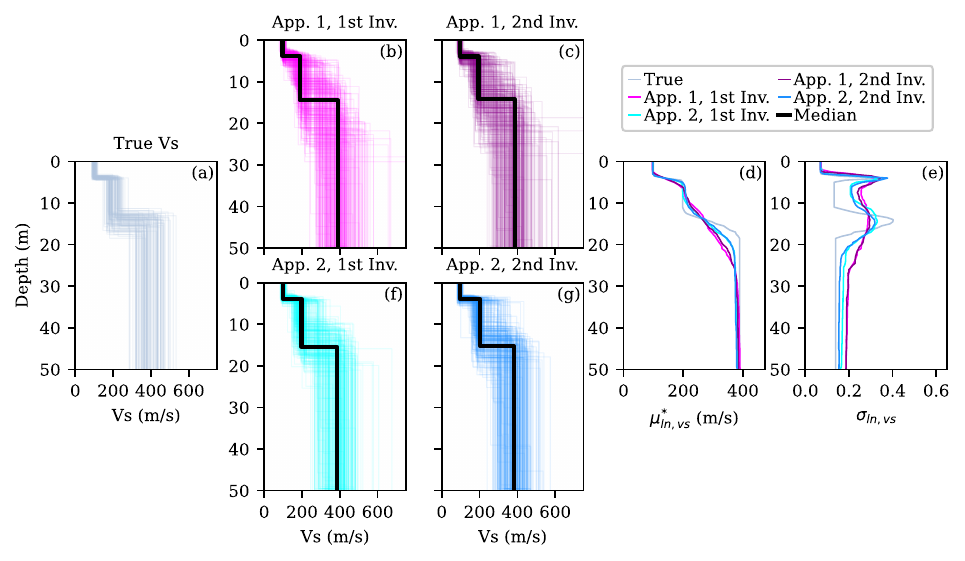}
	\caption{Summary of the shear wave velocity (Vs) profiles from the uncertainty-consistent inversions using Approach 1 and Approach 2 for the \textbf{three-layered model}. These include (a) 100 Vs profiles from the true distribution repeated from Figure \ref{fig:3}a, (b \& f) 250 uncertainty-consistent Vs profiles using Approach 1 and 2 following the 1st inversion with associated lognormal median profile, (c \& g) 250 uncertainty-consistent Vs profiles using Approach 1 and 2 following the 2nd inversion with associated lognormal median profile, (d) comparison of the discretized exponentiated lognormal median of Vs ($\mu_{ln,Vs}^*$) with depth, and (e) the discretized lognormal standard deviation of Vs with depth ($\sigma_{ln,Vs}$).}
	\label{fig:6}
\end{figure}

To evaluate the impact of estimating the correlation matrix on the uncertainty-consistent inversion results, Figure \ref{fig:6} compares the inverted Vs profiles to those initially sampled to develop the synthetic experimental dispersion data. If Approach 1 and 2 are successful in providing a reasonable estimate of the correlation matrix, then it is expected that the uncertainty-consistent Vs profiles would be consistent with the samples from the true distribution. Figure \ref{fig:6}a presents the 100 Vs profiles selected from the true Vs distribution as shown previously in Figure \ref{fig:3}a. The Vs profiles from the 1st (Figure \ref{fig:6}b and \ref{fig:6}c) and 2nd (Figures \ref{fig:6}f and \ref{fig:6}g) uncertainty-consistent inversions and their associated lognormal medians are shown for Approach 1 and 2, respectively. A qualitative comparison of Figure \ref{fig:6}b, \ref{fig:6}c, \ref{fig:6}f, and \ref{fig:6}g with Figure \ref{fig:6}a indicates that while the layering of the true Vs profiles is less clear in the suites of inverted Vs profiles, they are able to well capture the trend of Vs with depth and appropriately bound the associated Vs uncertainty. Furthermore, the layer-by-layer log-normal median Vs profiles for each suite are very similar to one another and clearly reflect the layering of the true profiles. To illustrate these more quantitatively, Figure \ref{fig:6}d and \ref{fig:6}e presents the discretized exponentiated lognormal median of Vs ($\mu_{ln,Vs}^*$) with depth, and the discretized lognormal standard deviation of Vs ($\sigma_{ln,Vs}$) with depth for all five sets of profiles. Examination of Figure \ref{fig:6}d and \ref{fig:6}e supports the qualitative observations, in particular, that the uncertainty-consistent inversions are appropriately capturing the true increase of Vs with depth (i.e., Figure \ref{fig:6}d) and the associated uncertainty (i.e., Figure \ref{fig:6}e). Furthermore, no significant differences are observed between the uncertainty-consistent Vs profiles derived after the 1st and 2nd inversion, therefore supporting the hypothesis that while the correlation matrix was shown to be improved (recall Figure \ref{fig:4}) this has a relatively minor effect on the resulting suite of Vs profiles.

As further quantitative support, Table \ref{table:1} presents the $\mu_{ln}^*$ and $\sigma_{ln}$ of each layer. The $\mu_{ln}^*$ values from the true Vs profiles and all four suites of Vs profiles are quite similar, indicating reasonable capture of the mean trend, whereas the $\sigma_{ln}$ values, while similar for all four suites of Vs profiles, tend to be larger than those for the true model, indicating the inverted profiles are able to bound the true uncertainty. We interpret the larger uncertainty values from the uncertainty-consistent inversions to be the result of inversion-derived uncertainty (i.e., uncertainty that arises through the solution of the inverse problem itself). We consider the primary contributor to inversion-derived uncertainty to be non-uniqueness in the inverse problem, that is, where multiple Vs profiles that may appear quite different map into similar locations in the dispersion space, thereby creating ambiguity when attempting to reverse the map from dispersion to Vs space. In addition to the Vs profiles themselves, it is also of interest to see potential effects on computed engineering parameters used as indicators for site effects. Herein, we consider two of the most common: the time-averaged shear wave velocity in the upper 30 m (Vs30) and the fundamental site resonant frequency ($f_{0}$) computed between 50 m (i.e., the base of the inverted profiles) and the ground surface. The $\mu_{ln}^*$ and $\sigma_{ln}$ of Vs30 and $f_{0}$ for the true and inverted Vs profiles are presented in Table \ref{table:2}. We observe strong consistency between the true and inverted values of Vs30 and $f_{0}$ across all results and do not observe significant differences between profiles extracted from the 1st and 2nd uncertainty-consistent inversion. In addition, the results from Table \ref{table:2} indicate that while the uncertainty-consistent Vs profiles in Figure \ref{fig:6} may visually appear highly variable, engineering features extracted from those profiles are consistent with the underlying distributions and tend to have comparable uncertainty. From these results, we infer that both Approach 1 and 2 are able to produce acceptable estimates of the correlation matrix and the derived uncertainty-consistent Vs profiles for a relatively simple three-layered subsurface model. Application of the uncertainty-consistent procedure to a more complex synthetic dataset may allow for insight into the relative contribution of inversion-derived uncertainty.

\begin{table}
\centering
\caption{Exponentiated lognormal median ($\mu_{ln}^*$) and lognormal standard deviation ($\sigma_{ln}$) of the time-averaged shear wave velocity in the upper 30 m (Vs30) and fundamental site frequency ($f_0$) computed between the base of the Vs profile (50 m) and the ground surface for the \textbf{three-layered model}. Results are presented for the true distribution as well as the suites of Vs profiles from Approach 1 and 2 after the 1st and 2nd uncertainty-consistent inversion.}
\label{table:2}
\resizebox{\textwidth}{!}{%
\begin{tabular}{ccccccccccc}
\hline
                               &                   &                   & \multicolumn{4}{c}{Approach 1}                                                  & \multicolumn{4}{c}{Approach 2}                                                  \\
                               & \multicolumn{2}{c}{True Distribution} & \multicolumn{2}{c}{Post 1st Inversion} & \multicolumn{2}{c}{Post 2nd Inversion} & \multicolumn{2}{c}{Post 1st Inversion} & \multicolumn{2}{c}{Post 2nd Inversion} \\
Parameter                      & $\mu_{ln}^*$      & $\sigma_{ln}$     & $\mu_{ln}^*$      & $\sigma_{ln}$      & $\mu_{ln}^*$      & $\sigma_{ln}$      & $\mu_{ln}^*$      & $\sigma_{ln}$      & $\mu_{ln}^*$      & $\sigma_{ln}$      \\ \hline
\multicolumn{1}{l}{Vs30 (m/s)} & 225               & 0.08              & 217               & 0.08               & 218               & 0.08               & 221               & 0.08               & 222               & 0.08               \\
\multicolumn{1}{l}{$f_0$ (Hz)} & 1.8               & 0.12              & 1.7               & 0.11               & 1.7               & 0.12               & 1.7               & 0.13               & 1.7               & 0.12               \\ \hline
\end{tabular}%
}
\end{table}

\subsection*{Complex Synthetic Dataset: Five-Layered Model}

\begin{figure}[t!]
    \centering
	\includegraphics[width=1.0\textwidth]{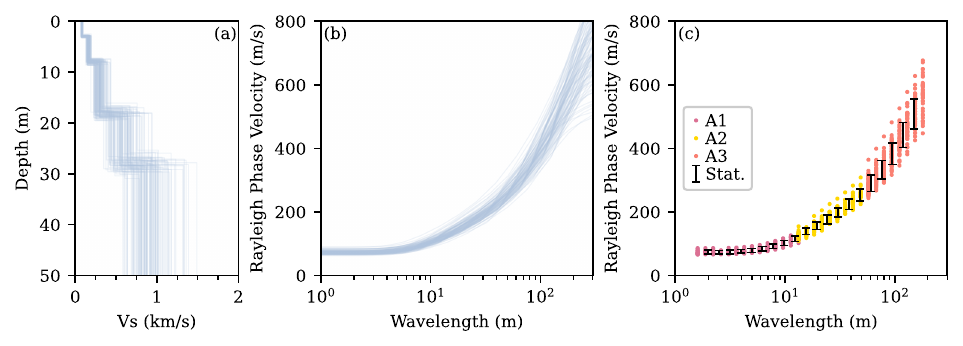}
	\caption{Creation of a synthetic experimental dispersion dataset for a \textbf{five-layered model} by first (a) generating 100 hypothetical shear wave velocity (Vs) profiles, (b) computing their corresponding theoretical dispersion data, and (c) defining the experimental dispersion data (i.e., sampled theoretical dispersion curves) over specific wavelength ranges as if they were measured by three arrays (A1, A2, and A3) of different size. The statistical representation of the synthetic experimental dispersion data is also presented in (c) with $\pm$ one standard deviation error bars.}
	\label{fig:7}
\end{figure}

To evaluate the generality of the conclusions drawn from the three-layered model considered in the prior section, this section considers a more-complex five-layered model. The synthetic experimental dispersion data was generated in a similar manner to that of the three-layered model. We again assume the H and Vs of each layer can be modeled as a multivariate-lognormal distribution, but now define the mean thickness and Vs of each layer to follow Profile 8 from the surface wave inversion benchmark dataset developed by Vantassel and Cox (\citeyear{vantassel_surface_2020}). We again define the lognormal standard deviation of Vs and H of each layer as well as the interparameter correlations following the same realism criteria described for the three-layered model. The $\mu_{ln}^*$ and $\sigma_{ln}$ of each layer's Vs and H are summarized in Table \ref{table:3}.  The 100 Vs profiles drawn from the multivariate lognormal distribution are shown in Figure \ref{fig:7}a. By defining Vp and $\rho$ for each layer in the same manner as before, the theoretical dispersion curves associated with the 100 Vs profiles were computed and are shown in Figure \ref{fig:7}b. Finally, the theoretical dispersion curves are divided into three groups (i.e., A1, A2, and A3) using the same procedure as for the three-layered model. The resulting synthetic experimental dispersion data and its corresponding $\pm$ one standard deviation dispersion statistics for the five-layered synthetic model are shown in Figure \ref{fig:7}c.

\begin{table}
\centering
\caption{Exponentiated lognormal median ($\mu_{ln}^*$) and lognormal standard deviation ($\sigma_{ln}$) of the parameters [i.e., layer-wise shear wave velocity (Vs) and thickness (H)] defining the inversion results of the synthetic \textbf{five-layered model}. Results are presented for the true distribution and for the distributions resulting from both Approach 1 and Approach 2 after the 1st and 2nd uncertainty-consistent inversion.}
\label{table:3}
\resizebox{\textwidth}{!}{%
\begin{tabular}{ccccccccccc}
\hline
          &                   &                   & \multicolumn{4}{c}{Approach 1}                                                  & \multicolumn{4}{c}{Approach 2}                                                  \\
          & \multicolumn{2}{c}{True Distribution} & \multicolumn{2}{c}{Post 1st Inversion} & \multicolumn{2}{c}{Post 2nd Inversion} & \multicolumn{2}{c}{Post 1st Inversion} & \multicolumn{2}{c}{Post 2nd Inversion} \\
Parameter & $\mu_{ln}^*$      & $\sigma_{ln}$     & $\mu_{ln}^*$      & $\sigma_{ln}$      & $\mu_{ln}^*$      & $\sigma_{ln}$      & $\mu_{ln}^*$      & $\sigma_{ln}$      & $\mu_{ln}^*$      & $\sigma_{ln}$      \\ \hline
Vs1 (m/s) & 80                & 0.05              & 80                & 0.05               & 80                & 0.04               & 80                & 0.04               & 80                & 0.04               \\
Vs2 (m/s) & 160               & 0.07              & 146               & 0.29               & 145               & 0.27               & 142               & 0.23               & 143               & 0.20               \\
Vs3 (m/s) & 300               & 0.07              & 228               & 0.37               & 221               & 0.34               & 223               & 0.32               & 224               & 0.32               \\
Vs4 (m/s) & 600               & 0.10              & 353               & 0.41               & 351               & 0.38               & 348               & 0.38               & 338               & 0.34               \\
Vs5 (m/s) & 900               & 0.15              & 810               & 0.32               & 807               & 0.31               & 836               & 0.30               & 796               & 0.28               \\
H1 (m)    & 3                 & 0.04              & 2.8               & 0.22               & 2.8               & 0.19               & 2.9               & 0.11               & 2.9               & 0.12               \\
H2 (m)    & 5                 & 0.04              & 3.5               & 0.71               & 3.4               & 0.71               & 3.1               & 0.74               & 3.3               & 0.67               \\
H3 (m)    & 10                & 0.05              & 3.2               & 0.98               & 3.4               & 0.95               & 3.1               & 0.96               & 3.4               & 0.87               \\
H4 (m)    & 10                & 0.06              & 7.0               & 1.15               & 6.8               & 0.99               & 8.8               & 1.05               & 8.8               & 0.94               \\ \hline
\end{tabular}%
}
\end{table}

\begin{figure}[t!]
    \centering
	\includegraphics[width=1.0\textwidth]{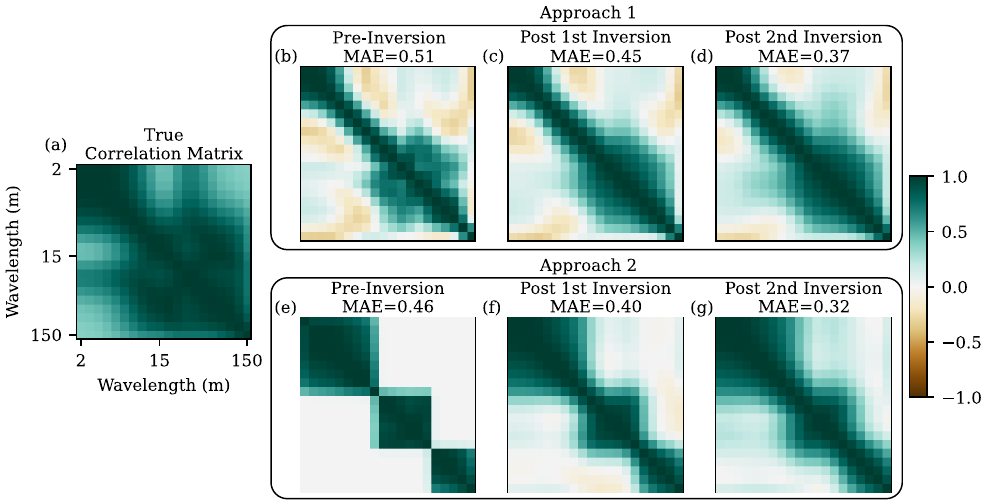}
	\caption{The correlation matrices estimated using Approach 1 and Approach 2 for the synthetic experimental dispersion dataset derived from a \textbf{five-layered model} compared to the (a) true correlation matrix. The correlation matrices for Approach 1 and 2 are shown (b \& e) before the 1st uncertainty-consistent inversion, (c \& f) after the 1st uncertainty-consistent inversion, and (d \& g) after the 2nd uncertainty-consistent inversion. The mean absolute error (MAE) between the estimated correlation matrix and the true correlation matrix is listed above each panel. All correlation matrices are plotted with the same orientation and scale as (a).}
	\label{fig:8}
\end{figure}

\begin{figure}[t!]
    \centering
	\includegraphics[width=0.5\textwidth]{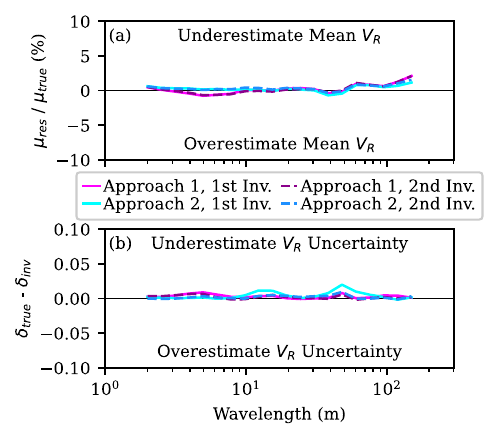}
	\caption{Evaluation of the uncertainty-consistent inversions of the synthetic experimental dispersion dataset derived from a \textbf{five-layered model} in terms of: (a) the phase velocity residual mean ($\mu_{res}$)[i.e., the difference between the mean of the true experimental dispersion data ($\mu_{true}$) and the mean of the inverted theoretical dispersion curves ($\mu_{inv}$)] normalized by $\mu_{true}$ and expressed in percent, (b) the phase velocity residual coefficient of variation [i.e., the difference between the coefficient of variation of the true experimental dispersion data ($\delta_{true}$) and the coefficient of variation of the inverted theoretical dispersion curves ($\delta_{inv}$)]. Results are shown for both Approach 1 and Approach 2 after the 1st (solid) and 2nd (dashed) uncertainty-consistent inversion.}
	\label{fig:9}
\end{figure}

As before, both Approach 1 and 2 are applied to the synthetic experimental dispersion data to develop a full surrogate dispersion data matrix, estimate the inter-wavelength correlations, and perform uncertainty-consistent inversions. Figure \ref{fig:8} compares the true correlation matrix in Figure \ref{fig:8}a with the estimated correlation matrices (\ref{fig:8}b \& \ref{fig:8}e) before the 1st uncertainty-consistent inversion, (\ref{fig:8}c \& \ref{fig:8}f) after the 1st uncertainty-consistent inversion, and (\ref{fig:8}d \& \ref{fig:8}g) after the 2nd uncertainty-consistent inversion. A quantitative comparison of the estimated correlation matrices indicates improvement following both the 1st and 2nd uncertainty-consistent inversion (i.e., reduced MAE). Note that for all of the inversions required, including the M1 inversion for Approach 1, a LN of 5 (i.e., LN=5) parameterization was used. Importantly, both Approaches 1 and 2 appear to be able to estimate a reasonable approximation of the true correlation matrix, despite having larger MAE values in comparison to the simpler, three-layered example. To check the uncertainty-consistent inversions are capturing the full multivariate normal distribution, Figure \ref{fig:9}a presents the difference between $\mu_{true}$ and $\mu_{inv}$ normalized by $\mu_{true}$ at each wavelength and Figure \ref{fig:9}b presents the difference between $\delta_{true}$ and $\delta_{inv}$ at each wavelength. Both indicate a good match between the statistics of the synthetic experimental dispersion data and the statistics of the theoretical dispersion data after uncertainty-consistent inversion. The percent difference between the true and inverted mean in Figure \ref{fig:9}a is generally within $\pm$ 2\%. The residual coefficient of variation in Figure \ref{fig:9}b is generally within $\pm$ 0.02 with no large deviations being observed. As noted previously in regards to the three-layered example, the absence of large deviations in the residual coefficient of variation for the five-layered example is believed to be the result of using a relatively-flexible parameterization that can well-fit each realization of the uncertainty-consistent inversion process. Based on these results, both Approach 1 and 2 appear to provide reasonable approximations of the correlation matrix for the five-layered model, with Approach 2 providing slightly lower MAE values. While there is a more pronounced trend toward decreasing the MAE between the 1st and 2nd uncertainty-consistent inversion, it is our opinion that the improvements are not significant enough to warrant the additional effort required to perform a 2nd uncertainty-consistent inversion to improve the estimate of the correlation matrix.

\begin{figure}[t!]
    \centering
	\includegraphics[width=1.0\textwidth]{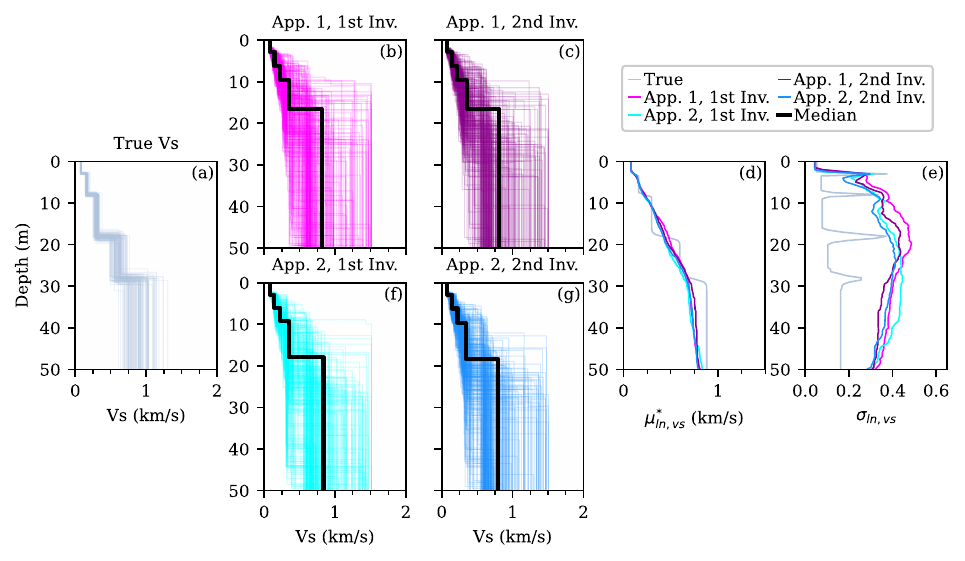}
	\caption{Summary of the shear wave velocity (Vs) profiles from the uncertainty-consistent inversions using Approach 1 and Approach 2 for the \textbf{five-layered model}. These include (a) 100 Vs profiles from the true distribution repeated from Figure \ref{fig:7}a, (b \& f) 250 uncertainty-consistent Vs profiles using Approach 1 and 2 following the 1st inversion with associated lognormal median profile, (c \& g) 250 uncertainty-consistent Vs profiles using Approach 1 and 2 following the 2nd inversion with associated lognormal median profile, (d) comparison of the discretized exponentiated lognormal median of Vs ($\mu_{ln,Vs}^*$) with depth, and (e) the discretized lognormal standard deviation of Vs with depth ($\sigma_{ln,Vs}$).}
	\label{fig:10}
\end{figure}

With the statistical representation of the synthetic experimental dispersion data having been well-captured by the uncertainty-consistent surface wave inversions, we present the corresponding Vs profiles. Figure \ref{fig:10}a presents the 100 Vs profiles generated from the true distribution of Vs, repeated from Figure \ref{fig:7}a, as reference for comparison with the Vs profiles from the uncertainty consistent inversions presented in Figure \ref{fig:10}b, \ref{fig:10}c, \ref{fig:10}f, and \ref{fig:10}g. While some outlier Vs profiles can be observed, the majority of the uncertainty-consistent Vs profiles qualitatively appear similar and exhibit an ability to capture the general trend of the true Vs distribution and the associated uncertainty, although they do not exactly capture the true model's deepest layer boundary at approximately 30 m. Nonetheless, the profiles are five layered, as shown by the lognormal median Vs profiles in Figure \ref{fig:10}b, \ref{fig:10}c, \ref{fig:10}f, and \ref{fig:10}g, and consistent with the true trend of Vs with depth. To examine the latter more quantitatively, Figure \ref{fig:10}d and \ref{fig:10}e presents $\mu_{ln,Vs}^*$ and $\sigma_{ln,Vs}$ with depth, respectively, for all five sets of profiles. In contrast to the results presented previously, a bias is observed between the true and inverted mean trend in Figure \ref{fig:10}d, with the inversion results tending to underestimate the Vs between approximately 30 and 45 m. This is believed to be the result of the relative insensitivity of the synthetic experimental dispersion data to the layer boundary observed in the true Vs profiles at a depth of approximately 30 m. This insensitivity of experimental dispersion data to deep contrasts relative to the longest wavelengths measured was observed in an earlier study by Vantassel and Cox (\citeyear{vantassel_swinvert_2021}), and unfortunately the use of an uncertainty-consistent inversion could not resolve the deep impedance contrast for reasons discussed later. In addition to the mean Vs trend, bias is also observed in the uncertainty of Vs with depth shown in Figure \ref{fig:10}e, with the $\sigma_{ln,Vs}$ of the inverted profiles exceeding that of the Vs profiles from the true distribution. As we discussed in regard to the three-layered profiles, the increase in Vs uncertainty in the inverted profiles is attributed to inversion-derived uncertainty, particularly non-uniqueness. The results from the three-layered profile in Figure \ref{fig:6}e indicate a notable, but still relatively minor, amount of inversion-derived uncertainty; however, results from the five-layered profile in Figure \ref{fig:10}e indicate a substantial increase in inversion-derived uncertainty. The inversion-derived uncertainty of the five-layered profile manifests itself primarily as layer boundary uncertainty. Experimental dispersion data is known to well-constrain the increase of Vs with depth, but not the exact placement of layer boundaries \citep{vantassel_swinvert_2021}. As a result, because $\sigma_{ln,Vs}$ mixes Vs uncertainty with layer-boundary uncertainty, the $\sigma_{ln,Vs}$ from profiles with more-uncertain boundaries are substantially higher than those without. These challenges highlight the need to incorporate other data, whenever possible, to reduce non-uniqueness (and thereby reduce inversion-derived uncertainty) on real projects. Examples of additional information that can be incorporated include knowledge of local geology, measurements of site resonant frequency from horizontal-to-vertical spectral ratio measurements, and site-specific invasive measurements (e.g., borehole logs).

Table \ref{table:3} presents the $\mu_{ln}^*$ and $\sigma_{ln}$ of the Vs and H of each layer of the suites of inverted profiles. Consistent with prior observations from Figure \ref{fig:10}, Table \ref{table:3} indicates that the inversion-derived Vs profiles from Approach 1 and 2, after both the 1st and 2nd uncertainty-consistent inversion, are similar. Furthermore, that the Vs of layers 3 and 4 and the H of layers 2 and 3 tend to be underestimated, resulting in profiles that tend to underestimate Vs in the 30 to 45 m range. The uncertainties of all model parameters are overestimated by the inversion-derived Vs profile suites, with the H of each layer being more-significantly over-estimated than Vs. Again, these results support that the increase in uncertainty of the inverted profiles is serving to quantify both the experimental-data-derived uncertainty and the inversion-derived uncertainty, resulting in a larger total measure of uncertainty in the model parameters. This is something that needs to be acknowledged more openly when presenting surface wave inversion results. However, while the uncertainty-consistent Vs profiles themselves exhibit relatively high uncertainty, the engineering parameters extracted from them, particularly Vs30 and $f_0$, are less influenced. Table \ref{table:4} shows that the estimates of Vs30 are practically unaffected, with a mean consistent with the true distribution but with a slightly larger uncertainty. The $f_0$ computed between 50 m and the ground surface is slightly affected with values slightly below the true resonance frequency (indicating softer structure on average, recall Figure \ref{fig:10}d) and with larger uncertainty. This serves to highlight that while the Vs profiles extracted using an uncertainty-consistent inversion may appear quite variable, they do maintain key site specific features by being constrained by the site-specific experimental dispersion data. The next section presents a real-data example and demonstrates the implications of uncertainty-consistent Vs profiles for use in site-specific ground response analyses.

\begin{table}
\centering
\caption{Exponentiated lognormal median ($\mu_{ln}^*$) and lognormal standard deviation ($\sigma_{ln}$) of the time-averaged shear wave velocity in the upper 30 m (Vs30) and fundamental site frequency ($f_0$) computed between the base of the Vs profile (50 m) and the ground surface for the \textbf{five-layered model}. Results are presented for the true distribution as well as the suites of Vs profiles from Approach 1 and 2 after the 1st and 2nd uncertainty-consistent inversion.}
\label{table:4}
\resizebox{\textwidth}{!}{%
\begin{tabular}{ccccccccccc}
\hline
           &                   &                   & \multicolumn{4}{c}{Approach 1}                                                  & \multicolumn{4}{c}{Approach 2}                                                  \\
           & \multicolumn{2}{c}{True Distribution} & \multicolumn{2}{c}{Post 1st Inversion} & \multicolumn{2}{c}{Post 2nd Inversion} & \multicolumn{2}{c}{Post 1st Inversion} & \multicolumn{2}{c}{Post 2nd Inversion} \\
Parameter  & $\mu_{ln}^*$      & $\sigma_{ln}$     & $\mu_{ln}^*$      & $\sigma_{ln}$      & $\mu_{ln}^*$      & $\sigma_{ln}$      & $\mu_{ln}^*$      & $\sigma_{ln}$      & $\mu_{ln}^*$      & $\sigma_{ln}$      \\ \hline
Vs30 (m/s) & 246               & 0.04              & 246               & 0.09               & 243               & 0.08               & 240               & 0.08               & 241               & 0.07               \\
$f_0$ (Hz) & 2.8               & 0.04              & 2.5               & 0.14               & 2.6               & 0.11               & 2.5               & 0.16               & 2.5               & 0.12               \\ \hline
\end{tabular}%
}
\end{table}

\section*{Real Data Example: Garner Valley Downhole Array Site}

From the synthetic data examples, Approaches 1 and 2 both appear to be viable solutions to the incomplete data matrix problem, thereby providing a means of applying the VC21 procedure to the case where multiple surface wave arrays of different size are deployed at the same site. Therefore, we apply Approaches 1 and 2 to a real surface wave dataset acquired at the GVDA site to develop, for the first time, uncertainty-consistent Vs profiles at a borehole array site.

The GVDA is located in southern California, approximately 70 miles northeast of San Diego and 90 miles southeast of Los Angeles. The GVDA exists in a seismically active region 7 km from the San Jacinto Fault and 35 km from the San Andreas Fault. Instrumentation at the GVDA began in 1989 and includes ground motion recording stations at the surface and at various depths \citep{archuleta_garner_1992}. The site is particularly interesting from an earthquake engineering perspective, as quite a few previous studies \citep{steidl_what_1996, bonilla_borehole_2002, teague_measured_2018, afshari_insights_2019, tao_insights_2019, vantassel_multi-reference-depth_2019} have attempted to match the frequency-dependent site amplifications observed at the GVDA using one-dimensional ground response analyses, but with limited success. Therefore, while not the primary objective of this work, we will compare the small-strain frequency-dependent site amplification estimated from the uncertainty-consistent Vs profiles developed using Approaches 1 and 2 to the empirical estimates of site amplification derived from surface and borehole recordings presented by Teague et al. (\citeyear{teague_measured_2018}).

\begin{figure}[t!]
    \centering
	\includegraphics[width=1.0\textwidth]{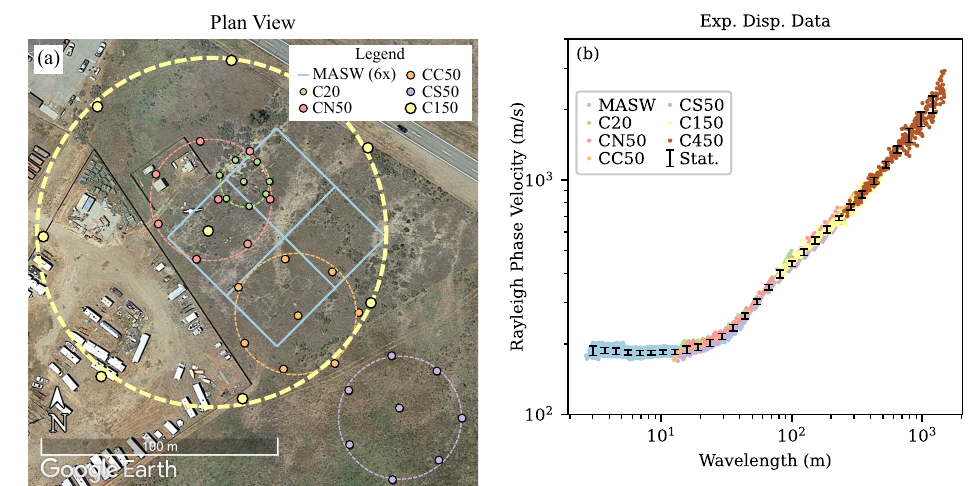}
	\caption{Experimental dataset collected at the Garner Valley Downhole Array (GVDA) site using (a) multiple active-source and passive-wavefield surface wave arrays and (b) extracted Rayleigh experimental dispersion data presented in terms of wavelength with $\pm$ one standard deviation error bars. In total, six active-source multichannel analysis of surface wave (MASW) arrays and six circular microtremor array measurements (MAM) of different diameters were deployed at the GVDA site. Note that the largest MAM array with a diameter of approximately 450 m (i.e., C450) is not shown in the plan view in panel (a), but its experimental dispersion data is shown in (b). }
	\label{fig:11}
\end{figure}

Experimental dispersion data was aggregated from two sources: active-source measurements by Vantassel et al. (\citeyear{vantassel_subsurface_2023}) and passive-wavefield measurements by Teague et al. (\citeyear{teague_measured_2018}). The active-source measurements include six lines of MASW; three running approximately NE to SW and three running in the perpendicular direction NW to SE. Figure \ref{fig:11}a indicates the location of each MASW array at the GVDA site. Each array was composed of 14, 3-component Fairfield nodal stations spaced at 5 m and buried just below the ground surface. Data was acquired using a sledgehammer and at least two source offsets between 5 and 17.5 m from either end of the array. Note that the MASW data utilized here is a small subset of a large dataset acquired at the GVDA site and subsequently published open-access \citep{vantassel_active-source_2023}. The reader is referred to Vantassel et al. (\citeyear{vantassel_subsurface_2023}) for details regarding the dataset. Experimental dispersion data was extracted from the MASW waveform recordings using frequency-domain beamformer \citep{zywicki_mitigation_2005} with cylindrical-wave steering vector and square-root-distance weighting, as implemented in the open-source software \emph{swprocess} \citep{vantassel_jpvantasselswprocess_2021}. Note that while three-component data was available for the active-source measurements, we only utilize the vertical component of the active-source recordings for extracting experimental Rayleigh wave dispersion data. The passive-wavefield measurements consist of six circular MAM arrays. The circular MAM arrays were composed of eight to ten Trillium compact 3-component broadband seismometers. The six circular arrays include: one approximately 20 m in diameter (C20), three approximately 50 m in diameter (CN50, CC50, CS50), one approximately 150 m in diameter (C150), and one approximately 450 m in diameter (C450). The positions of the MAM arrays at the GVDA site are shown in Figure \ref{fig:11}a, with the exception of C450, which was centered in the same area but is not shown due to its large scale.  The solid circular symbols associated with each MAM array indicate the positions where the broadband seismometer were buried, just below the ground surface, and the dashed circular lines indicate the approximate bounds of the array for visualization purposes. Experimental dispersion data was extracted from the passive-wavefield recordings using the Rayleigh three-component beamforming approach \citep{wathelet_rayleigh_2018}, as implemented in the open-source software Geopsy \citep{wathelet_geopsy_2020}. Following the extraction of the raw experimental dispersion data from the active-source and passive-wavefield recordings, the experimental dispersion data was interactively-trimmed and statistically combined following the workflow developed by Vantassel and Cox (\citeyear{vantassel_swprocess_2022}). The experimental dispersion data after interactive-trimming with summary $\pm$ one standard deviation statistics are presented in Figure \ref{fig:11}b. Through the use of multiple arrays, high-quality experimental dispersion data is observed over a wide range of wavelengths, between 3 and 1200 m.

\begin{figure}[t!]
    \centering
	\includegraphics[width=0.5\textwidth]{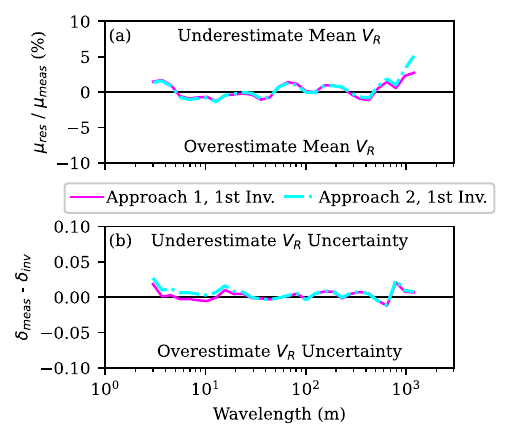}
	\caption{Evaluation of the uncertainty-consistent inversions at the Garner Valley Downhole Array (GVDA) site in terms of: (a) the phase velocity residual mean ($\mu_{res}$)[i.e., the difference between the mean of the measured experimental dispersion data ($\mu_{meas}$) and the mean of the inverted theoretical dispersion curves ($\mu_{inv}$)] normalized by $\mu_{meas}$ and expressed in percent, (b) the phase velocity residual coefficient of variation [i.e., the difference between the coefficient of variation of the measured experimental dispersion data ($\delta_{meas}$) and the coefficient of variation of the inverted theoretical dispersion curves ($\delta_{inv}$)].  Results are shown for both Approach 1 (solid) and Approach 2 (dash dot). }
	\label{fig:12}
\end{figure}

To solve the incomplete data matrix problem prior to uncertainty-consistent inversion, both Approach 1 and 2 were applied to the experimental dispersion data presented in Figure \ref{fig:11}b. For Approach 1, a relatively flexible nine-layered parameterization (i.e., LN=9) was selected for the M1 inversion to estimate the surrogate data matrix. Note the use of a LN=9 parameterization for the M1 inversion is somewhat arbitrary, however after observing no significant impacts on the inversion results (discussed later) we believe the direct selection of LN=9 rather than considering multiple parameterizations is a reasonable simplification. For Approach 2, the six active and passive surface wave arrays were reorganized based on wavelength range: short - including the MASW data, long - including the C450 MAM data, and intermediate - including the data from the remaining MAM arrays. Regrouping the data according to wavelength bandwidth rather than array type substantially reduced the number of combinations to be computed for Approach 2 and accelerated the estimation of the surrogate data matrix. Missing values in the surrogate data matrix from using Approach 2 were filled using variance-preserving imputation informed by the mean and standard deviation vectors estimated from the experimental dispersion data. Following the estimation of the inter-wavelength correlation matrices, uncertainty-consistent inversions were performed at the GVDA site using four different layering parameterization to account for epistemic uncertainty regarding the number of layers that best-represents the subsurface structure at the GVDA site. Note that different layering parameterizations were not used in the prior synthetic examples, as the true number of layers was assumed known to focus the discussion on the estimation of the inter-wavelength correlation matrix. In particular, for GVDA, we consider models with 5, 7, 9, and 12 layers (i.e., LN=5, LN=7, LN=9, and LN=12, respectively). The uncertainty-consistent inversion was repeated independently for each parameterization, with 250 realizations of the experimental dispersion data per parameterization. The sum result of the uncertainty-consistent inversions are a suite of 1000 ground models that capture the uncertainty in the experimental dispersion data, layering parameterization, and inversion itself (i.e., inversion-derived uncertainty). To demonstrate that the resulting 1000 ground models capture the experimental dispersion data's uncertainty, Figure \ref{fig:12} presents the percent difference in the mean and residual coefficient of variation between the statistical representation of the experimental dispersion data and the statistical representation of the inversion-derived theoretical dispersion data at all wavelengths. In general, the percent difference in the mean is within $\pm$ 2\% and the residual coefficient of variation is within $\pm$ 0.02. We do observe a slight underestimation of the mean (at most 5\%) at approximately a 1000 m wavelength. We attribute this to limiting the Vs in the inversion parameterizations to a maximum of 3500 m/s, a value consistent with crystalline granite, and thereby eliminating the possibility of getting unreasonably high Vs values inconsistent with the local geology. Negligible differences are observed between the results from Approach 1 and Approach 2, indicating that both approaches are reasonably capturing the statistics implied by the experimental dispersion data and can be inferred to be reasonable approximations of the surface wave dispersion trends governed by the GVDA site's subsurface structure.

\begin{figure}[t!]
    \centering
	\includegraphics[width=1.0\textwidth]{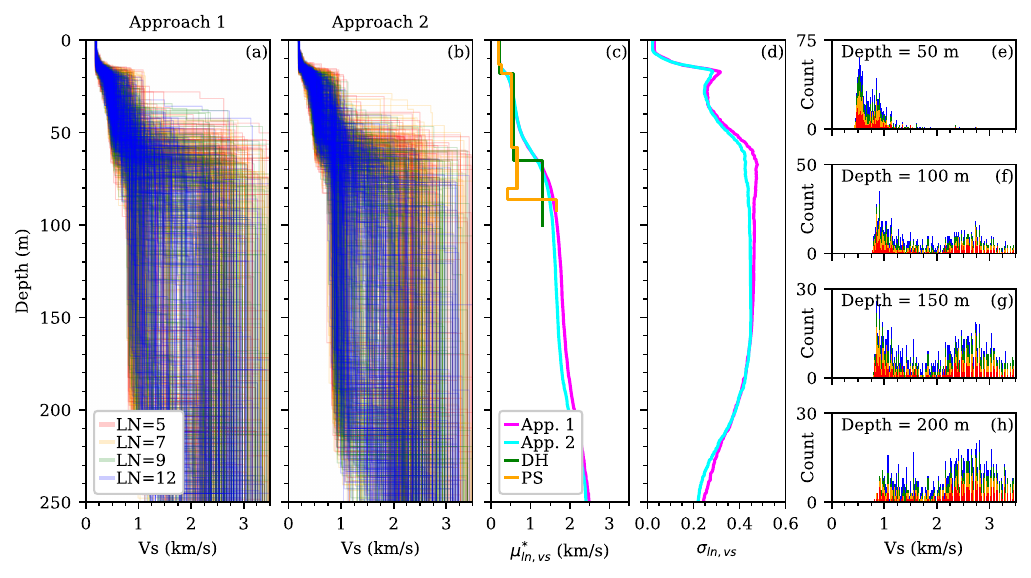}
	\caption{Uncertainty-consistent shear wave velocity (Vs) profiles developed at the Garner Valley Downhole Array (GVDA) site using (a) Approach 1 and (b) Approach 2. Shown in (a) and (b) are the best fit Vs profile to each of the 250 realizations of the experimental dispersion data for each of the four Layering by Number (LN) inversion parameterization. For each approach the (c) discretized exponentiated lognormal median of Vs ($\mu_{ln,Vs}^*$) and (d) discretized lognormal standard deviation of Vs ($\sigma_{ln,Vs}$) are shown with depth. For comparison, the invasive Vs profiles from downhole (DH) by Gibbs (\citeyear{gibbs_near-surface_1989}) and PS suspension logging (PS) by Steller (\citeyear{steller_new_1996}) are shown in panel (c). Histograms of Vs at (e) 50 m, (f) 100 m, (g) 150 m, and (h) 200 m indicate a bi-modal distribution below 50 m leading to larger than typical $\sigma_{ln,Vs}$.}
	\label{fig:13}
\end{figure}

Of particular importance are the 1000 Vs profiles associated with the 1000 uncertainty-consistent ground models aforementioned. Figure \ref{fig:13}a and \ref{fig:13}b present the 1000 uncertainty-consistent Vs profiles, 250 per layering parameterization, for Approach 1 and 2, respectively. The Vs profiles are qualitatively very similar between both approaches. Both sets of profiles are relatively well-constrained by the experimental dispersion data to a depth of approximately 70 m. At approximately 70 m, some of the profiles show a large impedance contrast, with Vs increasing from around 1000 m/s up to 3500 m/s (i.e., the maximum), whereas others maintain a Vs of approximately 1000 m/s until around 200 m depth, where Vs subsequently increases. The bi-modal distribution of the Vs profiles are shown with histograms of Vs at depths of 50 m, 100 m, 150 m, and 200 m in Figure \ref{fig:13}e – \ref{fig:13}h. Both Vs structures represent possible interpretations of the experimental dispersion data and demonstrate a consequence of epistemic uncertainty when performing surface wave inversion based solely on experimental dispersion data (i.e., without site-specific data). To better constrain the Vs structure and reduce epistemic uncertainty would require the use of a priori information, such as fundamental site frequency from horizontal-to-vertical spectral ratio measurements or deep borehole lithology logs, to inform the inversion parameterization. Alternatively, if insufficient a priori information was available to justify one interpretation over another, the data could be inverted twice with parameterizations consistent with the two interpretations and the results considered as separate branches of a logic tree. However, as the primary focus of this work is on the development of uncertainty-consistent Vs profiles rather than a detailed analysis of the data available at the GVDA site, we choose to focus here on examining the Vs profiles solely constrained by the experimental dispersion data, with the caveat that the variability observed is an upper bound of what could be expected if a more-detailed analysis was performed. To demonstrate the similarity between the profiles obtained using Approaches 1 and 2, $\mu_{ln,Vs}^*$ and $\sigma_{ln,Vs}$ are presented in Figure \ref{fig:13}c and \ref{fig:13}d, respectively. Both demonstrate similar trends, confirming qualitative observations made previously. In addition, invasive borehole measurements of Vs from downhole by Gibbs (\citeyear{gibbs_near-surface_1989}) and PS suspension logging by Steller (\citeyear{steller_new_1996}) are presented in Figure \ref{fig:13}c and compare quite favorably with $\mu_{ln,Vs}^*$. This supports observations made previously with the synthetic datasets, that while the Vs profiles themselves may appear highly variable the trend of the uncertainty-consistent Vs profiles tends to agree very well with the true subsurface structure. At depths less than approximately 15 m, $\sigma_{ln,Vs}$ are relatively low (\textless 0.2) compared to typical values assumed in the absence of site-specific measurements \citep{toro_probabilistic_1995, epri_seismic_2012, stewart_guidelines_2014}. However, at depths below approximately 60 m, because of the bi-modal distribution of Vs the $\sigma_{ln,Vs}$ is relatively high ($\approx$0.45), but if each mode of the distribution was handled separately, as you would if including the two interpretations on separate logic tree branches, $\sigma_{ln,Vs}$ reduces substantially ($\approx$0.25). Therefore, we conclude that Approaches 1 and 2 offer promising alternatives for solving the incomplete data-matrix problem for real field datasets and to enable the development of Vs profiles that account for epistemic and aleatory uncertainty from surface wave measurements. The ability of these Vs profiles to predict small-strain site response recorded at the GVDA site is discussed below.

\section*{Predicting Small-Strain Site Response at Garner Valley Downhole Array Site}

\begin{figure}[t!]
    \centering
	\includegraphics[width=1.0\textwidth]{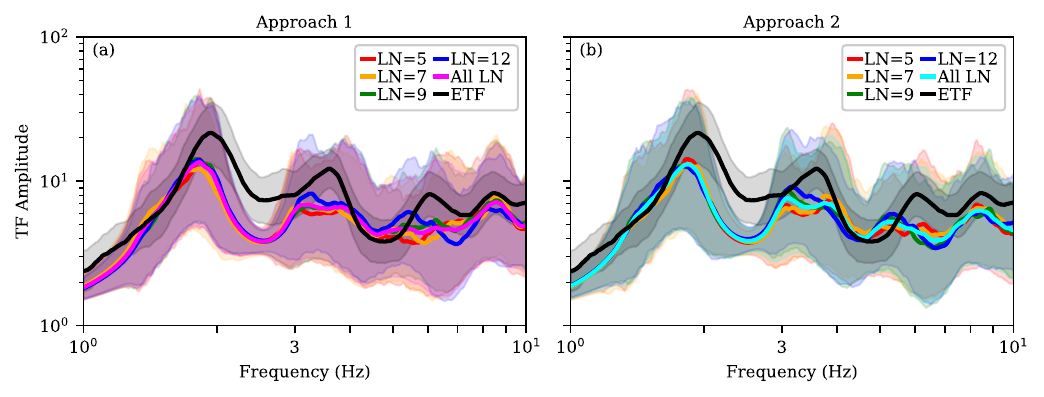}
	\caption{Exponentiated lognormal median empirical transfer function (ETF) and associated uncertainty ($\pm$ one standard deviation) at the Garner Valley Downhole Array (GVDA) site determined by Teague et al. (\citeyear{teague_measured_2018}) in comparison with the exponentiated lognormal median theoretical transfer functions (TTFs)  and associated uncertainty ($\pm$ one standard deviation) estimated from the uncertainty-consistent Vs profiles from each Layering by Number (LN) parameterization as well as the overall lognormal median TTF and associated uncertainty ($\pm$ one standard deviation) using Vs profiles from all parameterizations developed using (a) Approach 1 and (b) Approach 2. }
	\label{fig:14}
\end{figure}

To further evaluate the uncertainty-consistent Vs profiles developed, we compare the small-strain site amplification implied by the recovered Vs profiles to that observed from borehole and surface small-strain earthquake recordings at the GVDA site. In particular, we compare the lognormal median theoretical transfer function (TTF) computed from the 250 Vs profiles for each of the layering parameterizations, as well as across all parameterizations, for both Approaches 1 and 2 to the lognormal median empirical transfer function (ETF) developed by Teague et al. (\citeyear{teague_measured_2018}) at the GVDA site from 50 small-strain earthquake recordings (i.e., peak ground acceleration \textless 0.01 g). The TTFs were computed using the analytical solution for 1D resonance in a layered, damped elastic material assuming ``within'' conditions \citep{kramer_geotechnical_1996}. To define the required material properties for computing the TTF, mass density ($\rho$) and small-strain damping ($D_{min}$) were correlated from Vs for each layer using the relationships by Mayne (\citeyear{mayne_stress-strain-strength-flow_2001}) and Darendeli (\citeyear{darendeli_development_2001}), respectively. For the correlations, the material at the GVDA site was assumed to be non-plastic (plasticity index of zero), saturated below the ground water table (assumed at 5 m below the ground surface), and with an at-rest earth pressure coefficient ($K_0$) of 0.5. The TTFs (including $\pm$ one standard deviation bounds) for Approach 1 and 2 are compared with the ETF (including $\pm$ one standard deviation bounds) in Figure \ref{fig:14}a and \ref{fig:14}b, respectively. The lognormal median TTFs from Approach 1 and 2 are quite similar in terms of their agreement with the lognormal median ETF. As expected, the uncertainty associated with the TTFs is larger than that observed for the ETF, as the TTFs contain data-related as well as model-related uncertainties whereas the ETF only contain data-related uncertainty. Minimal variability (i.e., epistemic uncertainty) is observed between the TTFs developed from different LN parameterizations in terms of their lognormal median trend and uncertainty. This may indicate that when performing an uncertainty-consistent inversion the utilization of multiple parameterizations may not be as important when attempting to account for epistemic uncertainty, as observed previously \citep{cox_layering_2016, vantassel_swinvert_2021}. However, additional evidence is needed before a definitive conclusion can be made on whether or not the uncertainty-consistent inversions account for the combined effects of epistemic and aleatory variability. While not in perfect agreement, the median TTFs generally fit within the ETF's $\pm$ one standard deviation bounds, with the median TTFs having a tendency to underestimate the ETF (i.e., predict smaller amplification). The underestimation by the median TTFs, particularly in the vicinity of the site resonant frequency at 2 Hz, is attributed to the variability observed in the uncertainty-consistent Vs profiles below 70 m. The high variability below 70 m effectively averages out strong resonance effects at the site frequency, which helps to mitigate the significant over-estimation of small-strain site amplification at borehole array sites when using only a single Vs profile, as documented by many researchers (e.g., Thompson et al., \citeyear{thompson_taxonomy_2012}; Afshari and Stewart, \citeyear{afshari_insights_2019}; Tao and Rathje, \citeyear{tao_insights_2019}; Hallal et al., \citeyear{hallal_comparison_2022}). Nevertheless, underestimating site amplification is unconservative and it is possible that the subsurface variability present in the uncertainty-consistent Vs profiles could be reduced (and better estimates of site amplification developed as a consequence) by including additional site-specific information, such as site resonant frequency or geological layering from invasive measurements, into the uncertainty-consistent inversions. Regardless, the TTFs developed from the uncertainty-consistent Vs profiles qualitatively agree well with the ETF.

\begin{figure}[t!]
    \centering
	\includegraphics[width=0.5\textwidth]{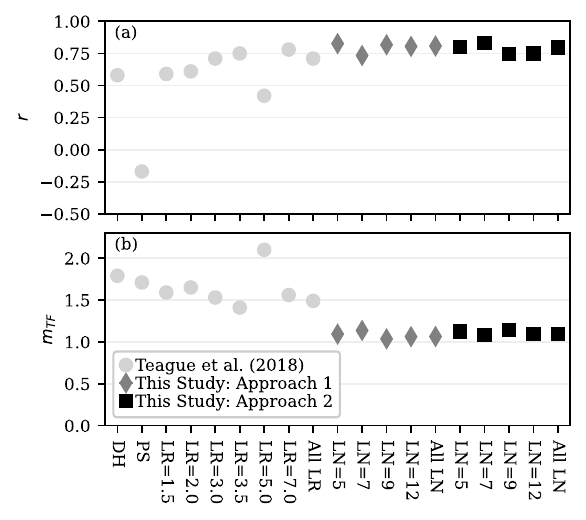}
	\caption{Quantitative goodness-of-fit between the theoretical and empirical transfer functions at the Garner Valley Downhole Array (GVDA) site using (a) Pearson correlation coefficient ($r$) and (b) transfer function misfit ($m_{TF}$). Values are shown for theoretical transfer functions developed using the Toro (\citeyear{toro_probabilistic_1995}) randomization procedure and base-case Vs profiles from downhole (DH) and PS suspension log (PS) testing, as reported by Teague et al. (\citeyear{teague_measured_2018}), the non-uncertainty-consistent surface wave inversion results using Layering Ratio (LR) parameterizations, as reported by Teague et al. (\citeyear{teague_measured_2018}), and the uncertainty-consistent surface wave inversion results using Layering by Number (LN) parameterizations performed using Approach 1 and 2 from the current study.}
	\label{fig:15}
\end{figure}

To evaluate the fit of the TTFs to the ETF more quantitatively, two statistical goodness-of-fit measures, Pearson correlation coefficient ($r$) and transfer function misfit ($m_{TF}$), are computed for each median TTF. For the sake of brevity, we refer the reader to Teague et al. (\citeyear{teague_measured_2018}) for the full mathematical definitions of the goodness-of-fit parameters, and instead here focus on the results. We do note that $r$ is a measure of correlation and, therefore, a value closer to one indicates better fit, whereas $m_{TF}$ is a measure of error and, therefore, a value closer to zero indicates better fit. The values of $r$ and $m_{TF}$ are relatively stable across the different LN parameterizations, with values of approximately 0.85 and 1.1, respectively, confirming quantitatively the observations made previously in regard to the relative insensitivity of the TTFs to the different inversion parameterizations and/or Approach 1 versus Approach 2. To provide context to the goodness-of-fit measures, Figure \ref{fig:15}a and \ref{fig:15}b summarize the $r$ and $m_{TF}$ values obtained in this study relative to those found by Teague et al. (\citeyear{teague_measured_2018}). In their study, Teague et al. (\citeyear{teague_measured_2018}) compared TTFs computed from the Toro-randomized base-case Vs profiles from downhole and PS suspension logging and those developed from traditional (i.e., non-uncertainty-consistent) surface wave inversion using different Layering Ratio (LR) parameterizations. Teague et al. (\citeyear{teague_measured_2018}) found that suites of Vs profiles from traditional surface wave inversion resulted in a better fit to the ETF, as shown by an increase in $r$ ($\approx$0.6 to $\approx$0.8) and a decrease in $m_{TF}$ ($\approx$1.75 to $\approx$1.5). In this context, we observe that by coupling the VC21 procedure with Approach 1 or Approach 2 developed through this study, uncertainty-consistent surface wave inversions are enabling substantially improved predictions (i.e., higher $r$ values and lower $m_{TF}$ values) than obtained in any previous study at the GVDA. This is strong evidence that the use of uncertainty-consistent inversions is a viable approach to account for Vs uncertainty in seismic hazard studies.

\section*{Conclusions}

The mapping of experimental data uncertainty acquired in the field to uncertainty in a site's Vs structure is a challenging but critical task required for site-specific probabilistic seismic hazard studies. Previously, Vantassel and Cox (\citeyear{vantassel_procedure_2021}) (i.e., VC21) proposed a procedure for performing this mapping during surface wave inversions. However, the VC21 procedure requires a full dispersion data matrix for computing inter-wavelength phase velocity correlations, which are not always possible to obtain directly from the experimental data, particularly when arrays of different sizes have been used in an attempt to measure broadband dispersion data and develop deep Vs profiles. Therefore, the present study evaluates two approaches (i.e., Approach 1 and Approach 2) to generalize the VC21 procedure by providing solutions to the incomplete data-matrix problem. Approach 1 uses a selection of theoretical dispersion curves from an initial, non-uncertainty-consistent global surface wave inversion to develop an estimate of the full data matrix. Approach 2 estimates the full data matrix by combining pieces of the data matrix obtained from the experimental dispersion measurements extracted from different surface wave arrays. Approach 1 is slower to perform than Approach 2 because it requires results from an initial, traditional surface wave inversion for its surrogate data matrix. However, it has the advantage that it can be applied to historic datasets where raw dispersion data is not available for each sub-array and only an estimate of the overall dispersion data's mean trend and associated uncertainty are provided. In contrast, Approach 2 does not require an initial inversion, but does require some additional statistical computations and access to the raw experimental dispersion data for each sub-array.

Both approaches have been presented and evaluated herein using two synthetic datasets and one real dataset. The two synthetic datasets include a relatively-simple dataset with a three-layered model and a more-complex dataset with a five-layered model. For both synthetic datasets, Approach 1 and Approach 2 were found to reasonably estimate the true correlation matrix and enable the recovery of uncertainty-consistent Vs profiles similar to the true Vs profiles. While the uncertainty of the recovered Vs profiles were higher than typically assumed ($\sigma_{ln,Vs}$) of approximately 0.2 and 0.4 for the three-layered and five-layered models, respectively), the engineering proxies computed from those Vs profiles, namely Vs30 and $f_0$, were able to be recovered with greater certainty. This highlights that while the uncertainty-consistent Vs profiles may appear highly variable they are effective at capturing a site's engineering behavior. As part of the evaluation of both approaches it was examined whether using the correlation matrix implied by the first uncertainty-consistent inversion could be used as input to a second uncertainty-consistent inversion to obtain more accurate results. While the estimate of the true correlation matrix did tend to improve after the second uncertainty-consistent inversion, no substantial improvement was observed in the capture of the experimental dispersion data's uncertainty, distribution of Vs profiles, or associated engineering parameters. As such, we conclude that given the effort required to perform a second uncertainty-consistent inversion and no significant improvement being observed in the resulting Vs profiles, a second uncertainty-consistent inversion is unjustified. Therefore, we recommend only one uncertainty-consistent inversion be performed when applying Approach 1 and Approach 2 in practice. Both approaches were applied to  a real dataset from the Garner Valley Downhole Array (GVDA) site, enabling for the first time the recovery of uncertainty-consistent Vs profiles at a borehole array site. The Vs profiles from Approach 1 and Approach 2 are consistent with Vs profiles from invasive borehole-based methods made previously at GVDA, and were found to be able to make better estimates of small-strain site amplification than achieved previously at the site. Based on these findings, the extended procedure for developing suites of uncertainty-consistent Vs profiles for use in probabilistic seismic hazard studies is shown to be a viable option for accounting for Vs uncertainty.

\section*{Acknowledgements}

The authors' thank Drs. David Teague and Andrew Stolte for their efforts to collect the microtremor array measurements at the GVDA site that enabled the development of uncertainty-consistent Vs profiles at the GVDA site. The authors would like to acknowledge Alan Yong, Robert Kent, Dr. Mohamad Hallal, and Dr. Mauro Aimar for assisting in the field data acquisition of the MASW data at the GVDA site. The full dataset acquired at the GVDA site, described in Vantassel et al. (\citeyear{vantassel_active-source_2023}), is published open-access on the DesignSafe-CI \citep{rathje_designsafe_2017} as Vantassel et al. (\citeyear{vantassel_subsurface_2023}). This work was partially supported by the U.S. National Science Foundation grants CMMI-1520808 and CMMI-1931162. However, any opinions, findings, and conclusions or recommendations expressed in this material are those of the authors and do not necessarily reflect the views of the National Science Foundation. The figures in this paper were created using Matplotlib 3.7.3 \citep{hunter_matplotlib_2007} and Inkscape 1.1.2.

\section*{CRediT Author Statement}

\textbf{Joseph P. Vantassel}: Conceptualization, Methodology, Software, Validation, Investigation, Writing - Original Draft, Writing - Review \& Editing, Visualization. \textbf{Brady R. Cox}: Conceptualization, Investigation, Writing - Review \& Editing, Funding acquisition

\bibliographystyle{plainnat}
\bibliography{ucvs2}

\end{document}